\documentclass[%
nofootinbib,
 amsmath,amssymb,
 aps, physrev,
prd,
twocolumn
]{revtex4-2}

\usepackage{graphicx}% Include figure files
\usepackage{dcolumn}% Align table columns on decimal point
\usepackage{bm}% bold math
\usepackage[colorlinks,linkcolor=magenta,citecolor=magenta,urlcolor=magenta]{hyperref}

\usepackage{bm,graphicx,dcolumn,epstopdf,epsf, latexsym,mathbbol, amssymb,amsmath,color,slashed, mathrsfs,mathcomp,simplewick}
\usepackage[center]{subfigure}
\usepackage{multirow}
\usepackage{graphicx}
\usepackage{makecell}
\usepackage{epstopdf}
\usepackage[table]{xcolor}
\newcommand{\orcid}[1]{\href{https://orcid.org/#1}{\includegraphics[width=10pt]{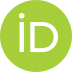}}}
\usepackage{booktabs} 
\graphicspath{{figs/}} % Specify the path to the figures folder

\begin{document}

\preprint{APS/123-QED}

\title{\textbf{Starobinsky like inflation and EGB Gravity in the light of ACT} 
}% 
\author{Yogesh\orcid{0000-0002-7638-3082}}%
 \email{yogesh@zjut.edu.cn, yogeshjjmi@gmail.com}
\affiliation{%
 Institute of Theoretical Physics \& Cosmology, School of Science, Zhejiang University of Technology, Hangzhou, China.
}%
%\orcid{0000-0003-1228-9107}
\author{Abolhassan Mohammadi\orcid{0000-0003-1228-9107}}
 \email{abolhassanm@zjut.edu.cn; gmail.com}
\affiliation{%
 Institute of Theoretical Physics \& Cosmology, School of Science, Zhejiang University of Technology, Hangzhou, China.
}
\author{Qiang Wu\orcid{0000-0002-3345-9905}}
\email{wuq@zjut.edu.cn}
\affiliation{%
 Institute of Theoretical Physics \& Cosmology, School of Science, Zhejiang University of Technology, Hangzhou, China.
}
\author{Tao Zhu\orcid{0000-0003-2286-9009}}
\email{zhut05@zjut.edu.cn}
\affiliation{%
 Institute of Theoretical Physics \& Cosmology, School of Science, Zhejiang University of Technology, Hangzhou, China.
}

% \author{Charlie Author}
% \homepage{http://www.Second.institution.edu/~Charlie.Author}
% \affiliation{
%  First affiliation for this author
% Anzhong_Wang@baylor.edu
% \affiliation{
%  second institution for this author
% }%
% \author{Delta Author}
% \affiliation{%
%  Authors' institution and/or address\\
%  This line break forced with \textbackslash\textbackslash
% }%

% \collaboration{CLEO Collaboration}%\noaffiliation

\date{\today}% It is always \today, today,
             %  but any date may be explicitly specified

\begin{abstract}
The ACT data shows an enhancement in the value of the scalar spectral index, as $n_s = 0.9743 \pm 0.0034$, leading to disfavoring many inflationary models, including the Starobinsky model. To satisfy the constraint made by ACT, we will investigate the Starobinsky potential within the Einstein-Gauss-Bonnet(EGB) gravity theory. EGB gravity is motivated by the higher-dimensional theory, which includes quadratic curvature correction terms and a coupling between the scalar field and the GB term that modifies the dynamical equations. The model is considered by using the slow-roll approximation method and the exact numerical approach for two different coupling functions. The results indicate that the model is in good agreement with the data and the results stand in $1\sigma$ of the ACT $r-n_s$ plane. Considering the running of the scalar spectral index also implies the consistency of the model with data. In addition, the parametric space of the free parameters of the EGB coupling is explored, where we find the acceptable region of the parameters in which the resulting $n_s$ and $r$ stay in $1\sigma$ of the ACT data. Next, the reheating phase is considered. It is determined that the model can simultaneously satisfy the constraint of ACT data and the reheating temperature constraints.
%It is determined that increasing the value of the coupling free parameters $\xi_2$ increases the scalar spectral index and the tensor-to-scalar ratio. For the $0.0012 < \xi_2 < 0.01$ range, the resulting values for $r-n_s$ stand in the $1\sigma$ region of P-ACT-BL-BK18. Then, it is concluded that the Starobinsky potential in the EGB gravity can satisfy the constraint of ACT data.  
\end{abstract}

%\keywords{Suggested keywords}%Use showkeys class option if keyword
                              %display desired
\maketitle

%\tableofcontents

%%%%%%%%%%%%%%%%%%%%%%%%%%%%%%%%%%%%%%%%%%%%%%%%%%%%%%%%%%%%%
%%%%%%%%%%%%%%%%%%%%%%%%%%%%%%%%%%%%%%%%%%%%%%%%%%%%%%%%%%%%%
%%%%%%%%%%%%%%%%%%%%%%%%%%%%%%%%%%%%%%%%%%%%%%%%%%%%%%%%%%%%%
%%%%%%%%%%%%%%%%%%%%%%%%%%%%%%%%%%%%%%%%%%%%%%%%%%%%%%%%%%%%%
\section{\label{introduction}
        Introduction}
Despite solving the hot big bang problem, the inflationary scenario predicts the primordial quantum perturbations, which are seeds for the large-scale structure of the universe~\cite{Guth:1980zm,Linde:1981mu,Mukhanov:1981xt,Sato:1981qmu,1996tyli.conf..771S,PhysRevLett.48.1220,Starobinsky:1982ee}. The scenario has received strong support from the observational data gathered over the last decades \cite{WMAP:2012nax, Planck:2015sxf,Planck:2018jri}, and has become the cornerstone of any cosmological model. The theory has been modified and studied in different ways and in different gravity theories \cite{Barenboim:2007ii,Franche:2010yj,
Fairbairn:2002yp,Mukohyama:2002cn,Feinstein:2002aj,Padmanabhan:2002cp,
Spalinski:2007dv,Bessada:2009pe,Weller:2011ey,
maeda2013stability,alexander2015dynamics,
maartens2000chaotic,Mohammadi:2020ake,Mohammadi:2020ctd,berera1995warm,Sheikhahmadi:2019gzs,Mohammadi:2018oku,Mohammadi:2019dpu,Mohammadi:2018zkf,Mohammadi:2019qeu,Mohammadi:2020ftb,Mohammadi:2021wde,Mohammadi:2021gvf,Mohammadi:2022vru,Mohammadi:2022fiv}. During the inflationary phase, the universe is dominated by a scalar field that slowly rolls down toward the minimum of the potential. Due to this slow rolling, the universe undergoes a quasi-de Sitter expansion and expands extremely in a very short time.  

The recent data released by the Atacama Cosmology Telescope (ACT) reveals that the value of the scalar spectral index $n_s$ is higher than the value reported by Planck. The value of $n_s$ is reported as $n_s = 0.9709 \pm 0.0038$ as a result of a joint analysis of Planck and ACT data \cite{ACT:2025fju, ACT:2025tim}. The value increases to $n_s = 0.9743 \pm 0.0034$ by incorporating further data from CMB, BAO, and DESI \cite{ACT:2025fju, ACT:2025tim}. This new observational result implies that some well-accepted inflationary models, such as the Starobinsky model, which was consistent with the previous data, stand on the $2\sigma$ boundary and become disfavored. The shift in the $r-n_s$ diagram emphasizes the importance of revising and refining the consistent inflationary models. Therefore, it is necessary to reconsider the broad class of inflationary models compared to the latest observational data\cite{Kallosh:2025rni,Aoki:2025wld,Dioguardi:2025vci,Salvio:2025izr,Brahma:2025dio,Gao:2025onc,Drees:2025ngb,Zharov:2025evb,Yin:2025rrs,Liu:2025qca,Gialamas:2025ofz}.

The universe is in the very cold stage at the end of inflation, and a reheating phase is required to warm it up\cite{Kofman:1994rk}. During the reheating phase, the energy stored in the scalar field decays, creating standard particles. The produced particles interact with each other, and the universe warms up, providing a smooth transition to the radiation phase (refer to  \cite{Lozanov:2019jxc, Kofman:1997pt} for a more detailed explanation about the reheating process). There is no strict observational constraint on the reheating temperature. There is a lower range for the reheating temperature originated from the BBN as $T_{\rm BBN} \simeq 10^{-2} \; {\rm GeV}$, and an upper limit coming from the scale of inflation, as $T_{re} < 10^{16} \; {\rm GeV}$. Then, there is a wide valid range for the reheating temperature.

Among different cosmological models, those with a non-minimal coupling provide tools to get low values of the tensor-to-scalar ratio, $r$~\cite{Bezrukov:2007ep,Bezrukov:2010jz}. These models have received interest due to the small upper bound on the tensor-to-scalar ratio as $r < 0.038$ (P-ACT-LB-BK18). On the other hand, the gravity theories with higher order of curvature term, which are motivated by superstring theory, determine that besides the higher order of the curvature terms, there should be a non-minimal coupling between the geometry and the scalar field. Therefore, there is a strong motivation for considering modified gravity theories to understand the evolution of the universe in the very early times. It is realized that the modified gravity theories can reconcile even the potentials disfavored in the standard gravity~\cite{Kallosh:2013pby,1982PhLB..117..175S,Starobinsky:1983zz,Barvinsky:1994hx,Cervantes-Cota:1995ehs,Bezrukov:2007ep,Barvinsky:2008ia,DeSimone:2008ei,Gialamas:2020vto,Bezrukov:2008ej,Barvinsky:2009ii,Bezrukov:2010jz,Rubio:2018ogq,Koshelev:2020xby,Gialamas:2023flv,Gialamas:2022xtt,Gialamas:2021enw,Gialamas:2020snr,Kim:2025ikw,Kim:2025dyi}. Here, we are interested in exploring the EGB gravity, a modified gravity theory inspired by the higher-dimensional theory, including quadratic curvature corrections. The theory contains a scalar field with a non-minimal coupling to the GB term described by $\xi(\phi)$~\cite{vandeBruck:2015gjd,Guo:2009uk,Guo:2010jr,Koh:2016abf,Pozdeeva:2020shl,Satoh:2008ck,Jiang:2013gza,Koh:2014bka,Koh:2018qcy,Mathew:2016anx,Pozdeeva:2020apf,Pozdeeva:2016cja,Nozari:2017rta,Yi:2018dhl,Odintsov:2018zhw,Fomin:2019yls,Fomin:2020hfh,Kleidis:2019ywv,Odintsov:2020sqy,Odintsov:2020zkl,Kawai:2021bye,Kawai:2017kqt,Oikonomou:2022xoq,Oikonomou:2022ksx,Cognola:2006sp,Odintsov:2020xji,Odintsov:2020mkz,Oikonomou:2020sij,Nojiri:2019dwl,Fomin:2019yls,Ashrafzadeh:2023ndt,Oikonomou:2024etl,Oikonomou:2024jqv,Odintsov:2023weg,Kawai:1999pw,Kawai:1998ab,Nojiri:2024hau,Nojiri:2024zab,Elizalde:2023rds,Nojiri:2023mvi,Odintsov:2023aaw,Odintsov:2022rok,Odintsov:2022rok,Odintsov:2021urx,Kawai:2017kqt,Kawai:2023nqs,Mudrunka:2025xcg}. The Gauss-Bonnet term plays an essential role in determining the dynamics of the universe. It acts as a quantum correction to the Einstein-Hilbert action in string theory.   

Here, we investigate inflation in EGB gravity frame with two GB coupling functions as ``tanh"~\cite{Kawai:2021edk,Ashrafzadeh:2023ndt,Yogesh:2025hll} and ``exp"~\cite{Pozdeeva:2020shl,Jiang:2013gza,Yi:2018gse,Odintsov:2018zhw,Kleidis:2019ywv,Rashidi:2020wwg}. The potential of the scalar field is assumed to be described by the Starobinsky potential~\cite{Starobinsky:1980te}. Although the potential has been disfavored in the standard gravity, we determine that the predicted values of the scalar spectral index and the tensor-to-scalar ratio stay within the $1\sigma$ region, and the model exhibits good consistency with the ACT data. Next, we consider the reheating phase and the behavior of the reheating e-fold and reheating temperature versus the scalar spectral index. It is found that the model can come to good agreement with the ACT data, and simultaneously satisfy the reheating temperature constraint for $\omega_{re} > 1/3$.

The paper is organized as follows: in Sec.\ref{sec:model}, we briefly review the EGB gravity theory. Then, we discuss the inflationary and reheating phases in Sec.\ref{sec:inflation} and rewrite the main dynamical equations by applying the slow-roll approximations. The results of the model from the slow-roll approach and the exact numerical approach are compared to the ACT data in Sec.\ref{sec:result}. Finally, we summarize the results in Sec.\ref{sec:conclusion}. \\

%%%%%%%%%%%%%%%%%%%%%%%%%%%%%%%%%%%%%%%%%%%%%%%%%%%%%%%%%%%%%
%%%%%%%%%%%%%%%%%%%%%%%%%%%%%%%%%%%%%%%%%%%%%%%%%%%%%%%%%%%%%
%%%%%%%%%%%%%%%%%%%%%%%%%%%%%%%%%%%%%%%%%%%%%%%%%%%%%%%%%%%%%
%%%%%%%%%%%%%%%%%%%%%%%%%%%%%%%%%%%%%%%%%%%%%%%%%%%%%%%%%%%%%
%%%%%%%%%%%%%%%%%%%%%%%%%%%%%%%%%%%%%%%%%%%%%%%%%%%%%%%%%%%%%
%%%%%%%%%%%%%%%%%%%%%%%%%%%%%%%%%%%%%%%%%%%%%%%%%%%%%%%%%%%%%
\section{\label{sec:model}
        EGB gravity}
The action of the model is given by~\cite{Pozdeeva:2020apf}
\begin{equation}
\label{action1}
S=\int d^4x\sqrt{-g}\left[ \frac{R}{2} - \frac{1}{2}g^{\mu\nu}\partial_\mu\phi\partial_\nu\phi-V(\phi)-\frac{1}{2}\xi(\phi){\cal G}\right],
\end{equation}

where $g$ is the determinant of the metric $g_{\mu\nu}$, $R$ is the Ricci scalar constructed based on the metric $g_{\mu\nu}$, and $V(\phi)$ is the potential of the scalar field $\phi$. The term $\mathcal{G}$ carries the quadratic curvature correction, given as $\mathcal{G}=R_{\mu\nu\rho\sigma}R^{\mu\nu\rho\sigma}-4R_{\mu\nu}R^{\mu\nu}+R^2$. 
% \begin{equation}
%     \mathcal{G}=R_{\mu\nu\rho\sigma}R^{\mu\nu\rho\sigma}-4R_{\mu\nu}R^{\mu\nu}+R^2.
% \end{equation}
Assuming that the geometry of the universe is described by a spatially flat FLRW metric, the modified Friedmann equations are obtained as
% Assuming that the geometry of the universe is described by a spatially flat FLRW metric, 
% \begin{equation}
%     ds^2 = -dt^2 + a^2(t) \big( dx^2 + dy^2 + dz^2 \big),
% \end{equation}
% where $a(t)$ is the scale factor of the universe, the modified Friedmann equations are given as
\begin{eqnarray}
6H^2\left(1-4\xi_{,\phi} \dot\phi H\right)&=&\dot\phi^2+2V,
\label{Equ00} \\
2\dot H\left(1 - 4\xi_{,\phi}\dot\phi H\right)&=&{}-\dot\phi^2+4H^2 \; \Psi,
\label{Equ11}
\end{eqnarray}
and for the field equation, we have
\begin{equation}
\ddot\phi+3H\dot\phi = {} -V_{,\phi} -12H^2\xi_{,\phi}\left(\dot{H}+H^2\right),\label{Equphi}
\end{equation}
here, $\Psi$ is defined as $\Psi = \xi_{,\phi\phi}\dot\phi^2+\xi_{,\phi}\ddot\phi-H\xi_{,\phi}\dot\phi$, $H = \dot{a}/a$ is the Hubble parameter, and the dot stands for the derivative with respect to cosmic time. In the case of varying coupling function, i.e. $\xi_{,\phi} \neq 0$, the equations \eqref{Equ11} and \eqref{Equphi} are not the dynamical equations. After working with Eqs.~(\ref{Equ00}) and (\ref{Equphi}) and doing some algebra, the dynamical equation can be obtained as (refer to \cite{Pozdeeva:2024ihc} for more details)
\begin{widetext}
\begin{equation}
\label{DynSYSN}
\begin{split}
\frac{d\phi}{dN}=&\,\chi,\\
\frac{d\chi}{dN}=&\,\frac{1}{H^2\left(B-2\xi_{,\phi}H^2\chi\right)} \Big(
3\left[3-4\xi_{,\phi\phi} H^2\right]\xi_{,\phi}H^4\chi^2+ \left[3B+2\xi_{,\phi}V_{,\phi}-3\right]H^2\chi-4V^2 X \Big) -\frac{\chi}{2H^2}\frac{dH^2}{dN},\\
\frac{dH^2}{dN}=&\,\frac{H^2}{2\left(B-2\xi_{,\phi}H^2\chi\right)} \Big( \left(4\xi_{,\phi\phi}H^2-1\right)\chi^2-16\xi_{,\phi}H^2\chi-16 V^2 \xi_{,\phi}X \Big).
\end{split}
\end{equation}  
\end{widetext}
where $\chi$ is defined as $\chi = \dot\phi / H$, and we applied a variable change to derive the equations in terms of the number of e-folds $N = \ln\big( a / a_e \big)$ than the cosmic time $t$, using $dN = H dt$. The quantities $A$ and $X$ are defined as
\begin{equation}
B=12\xi_{,\phi}^2H^4 + \frac{1}{2},\qquad
X=\frac{1}{4V^2}\left(12\xi_{,\phi} H^4+V_{,\phi} \right).
\end{equation}

% \begin{equation}
% \label{DynSYS}
% \begin{split}
% \dot\phi=&\psi,\\~~~~~~~~~~~~~
% \dot\psi=&\frac{1}{B-2\xi_{,\phi}H\psi}\left\{
% 3\left[3-4\,\xi_{,\phi\phi} H^2 \right]\xi_{,\phi}H^2\psi^2+\left[3B+2\,\xi_{,\phi}V_{,\phi}-6U_0\right]H\psi-\frac{V^2}{U_0}X\right\},\\
% \dot H=&\frac{1}{4\left(B-2\,\xi_{,\phi}H\psi\right)}\left\{\left(4\,\xi_{,\phi\phi}H^2-1\right)\psi^2-16\xi_{,\phi}H^3\psi-4\frac{V^2}{U_0^2}\,\xi_{,\phi} H^2 X\right\},
% \end{split}
% \end{equation}\\
% where the used quantities $B$ and $X$ are defined as
% \begin{equation}
% B=12\xi_{,\phi}^2H^4+U_0,\qquad
% X=\frac{U_0^2}{V^2}\left(12\xi_{,\phi} H^4+V_{,\phi} \right).
% \end{equation}
% To consider the inflationary phase, it is a better choice to work with the number of e-fold, $N = \ln\big( a / a_e \big)$ than the cosmic time. Doing a variable change, $dN = H dt$, the dynamical equations are rewritten as 
% \begin{equation}
% \label{DynSYSN}
% \begin{split}
% \frac{d\phi}{dN}=&\,\chi,\\
% \frac{d\chi}{dN}=&\,\frac{1}{H^2\left(B-2\xi_{,\phi}H^2\chi\right)}\left\{
% 3\left[3-4\xi_{,\phi\phi} H^2\right]\xi_{,\phi}H^4\chi^2+ \left[3B+2\xi_{,\phi}V_{,\phi}-6U_0\right]H^2\chi-\frac{V^2}{U_0}X\right\}\\
% &{}-\frac{\chi}{2H^2}\frac{dH^2}{dN},\\
% \frac{dH^2}{dN}=&\,\frac{H^2}{2\left(B-2\xi_{,\phi}H^2\chi\right)}\left\{\left(4\xi_{,\phi\phi}H^2-1\right)\chi^2-16\xi_{,\phi}H^2\chi-4\frac{V^2}{U_0^2}\xi_{,\phi}X\right\}.
% \end{split}
% \end{equation}
% in which we have defined $Q\equiv H^2$ and $\chi=\psi/H$. 

The inflationary phase is usually described by the slow-roll parameters, which are defined as 
\begin{equation}
\label{epsilon}
\varepsilon_1 ={}-\frac{\dot{H}}{H^2}={}-\frac{d\ln(H)}{dN},\qquad \varepsilon_{i+1}= \frac{d\ln|\varepsilon_i|}{dN},\quad i\geqslant 1,
\end{equation}
In addition to these parameters, in EGB gravity, we have some other slow-roll parameters which are related to the coupling term as
\begin{equation}
\label{delta}
\delta_1= %\frac{2}{U_0}\xi_{,\phi}H\dot\phi=
4\xi_{,\phi}H^2\chi,\qquad \delta_{i+1}=\frac{d\ln|\delta_i|}{dN}, \quad i\geqslant 1.
\end{equation}
The evolution equations can be rewritten in terms of the slow-roll parameters explained in \cite{Pozdeeva:2024ihc}. Then, the scalar spectral index and the tensor-to-scalar ratio are also expressed in terms of these slow-roll parameters as
\begin{equation}
\label{ns_slr}
n_s=1-2\varepsilon_1-\frac{2\varepsilon_1\varepsilon_2-\delta_1\delta_2}{2\varepsilon_1-\delta_1}, \qquad r=8|2\varepsilon_1-\delta_1|. 
%=1-2\varepsilon_1-\frac{d\ln(r)}{dN}=1+\frac{d}{dN}\ln\left(\frac{Q}{U_0r}\right),
\end{equation}
% \begin{equation}
% \label{r_slr}
% r=8|2\varepsilon_1-\delta_1|.
% \end{equation}

%%%%%%%%%%%%%%%%%%%%%%%%%%%%%%%%%%%%%%%%%%%%%%%%%%%%%%%%%%%%%
%%%%%%%%%%%%%%%%%%%%%%%%%%%%%%%%%%%%%%%%%%%%%%%%%%%%%%%%%%%%%
%%%%%%%%%%%%%%%%%%%%%%%%%%%%%%%%%%%%%%%%%%%%%%%%%%%%%%%%%%%%%
%%%%%%%%%%%%%%%%%%%%%%%%%%%%%%%%%%%%%%%%%%%%%%%%%%%%%%%%%%%%%
%%%%%%%%%%%%%%%%%%%%%%%%%%%%%%%%%%%%%%%%%%%%%%%%%%%%%%%%%%%%%
%%%%%%%%%%%%%%%%%%%%%%%%%%%%%%%%%%%%%%%%%%%%%%%%%%%%%%%%%%%%%
\section{\label{sec:inflation}
        Inflation and reheating phases}
This section considers the inflationary and reheating phases within the introduced gravity frame. The potential of the scalar field is taken to be the Starobinsky potential, which has been disfavored in the standard gravity. The potential is given as~\cite{Starobinsky:1980te}\footnote{The potential introduced in Eq.\eqref{potential} appears in the $R^2$ model, non-minimal coupling model, as well as the alpha-attractor model.}
\begin{equation}\label{potential}
    V(\phi) = V_0 \; \left( 1 - e^{- \sqrt{\frac{2}{3}} \phi}  \; \right)^2
\end{equation}
We consider two different commonly used GB couplings as \cite{Khan:2022odn,Gangopadhyay:2022vgh,Yogesh:2024mpa,Pozdeeva:2020shl,Jiang:2013gza,Yi:2018gse,Odintsov:2018zhw,Kleidis:2019ywv,Rashidi:2020wwg}
\begin{equation}\label{coupling_function}
    \xi(\phi) = \frac{\xi_1}{V_0} \;  \tanh(\xi_2 \; \phi), \quad 
    \xi(\phi) = \frac{\xi_1}{V_0} \; e^{-\xi_2 \; \phi}
\end{equation}
where $\xi_1$ and $\xi_2$ are two constant parameters. The constant $V_0$ is defined here to simplify future calculations, and it does not affect the generality of the consideration. We follow the effective potential method, which has been introduced in \cite{Pozdeeva:2019agu,Pozdeeva:2020apf,Vernov:2021hxo}, where the stability of de Sitter solutions of the model has been studied. The effective potential is given in terms of the potential and the coupling term as  
\begin{equation}
\label{Veff}
V_{eff}(\phi)={}-\frac{1}{4V(\phi)}+\frac{1}{3}\xi(\phi).
\end{equation}

To consider the result of the model, we are going with a numerical approach and the slow-roll method. %The last two approaches were introduced in \cite{Pozdeeva:2024ihc} and were used in [??] for considering the inflationary phase. 
In the standard slow-roll approximation, the slow-roll parameters, introduced by Eqs.\eqref{epsilon} and \eqref{delta}, are assumed to be small; known as the slow-roll approximations. %Due to this approximation, the dynamical equations are simplified and easier to solve. 
% In the new approximations I and II ........................ The dynamical equations for the new slow-roll approximations I and II have been considered in detail, and we will not repeat them. ................ \\ 
By applying the slow-roll approximations, the dynamical equations are simplified, and one can find
\begin{equation}
\label{N1}
\frac{dN}{d\phi}\simeq{}-\frac{1}{4V{V_{eff}}_{,\phi}}\,.
\end{equation}
Solving the integral, the scalar field is obtained in terms of the number of e-folds. In addition, the Hubble parameter and $\chi$ are obtained in terms of the potential and coupling term as
\begin{equation}\label{EquSsr}
H^2 \simeq \frac{V}{3}, \qquad 
\chi\simeq {}-4{V_{eff}}_{,\phi}.
\end{equation}
Applying the approximations, the slow-roll parameters are obtained in terms of the scalar field. Substituting the obtained solution form Eq.\eqref{N1} in the slow-roll parameters, the scalar spectral index and the tensor-to-scalar ratio are resulted from Eq.\eqref{ns_slr}. \\

After inflation, the universe will be repopulated due to the energy decay from the scalar field to other standard particle fields, and it will enter a hot thermal bath filled with relativistic particles. Due to the lack of a direct observational constraint for reheating, exploring the phase indirectly by analysing the reheating temperature using the inflationary observable parameters is beneficial \cite{Cook:2015vqa}. Considering that the equation of state parameter of the matter in the reheating phase could be described by the constant parameter $\omega_{re}$, one arrives at \cite{Cook:2015vqa,Khan:2022odn,Gangopadhyay:2022vgh,Adhikari:2019uaw}.
\begin{equation}
N_{re}= \frac{4}{ (1-3w_{re} )}   \left[61.488  - \ln \left(\frac{ V_{end}^{\frac{1}{4}}}{ H_{k} } \right)  - N_{k}   \right]
\label{Nre}
\end{equation}
\begin{equation}
T_{re}= \left[ \left(\frac{43}{11 g_{re}} \right)^{\frac{1}{3}}    \frac{a_0 T_0}{k_{}} H_{k} e^{- N_{k}} \left[\frac{3^2 \cdot 5 V_{end}}{\pi^2 g_{re}} \right]^{\beta}  \right]^\gamma,
\label{Tre}
\end{equation} 
where $T_{re}$ and $N_{re}$ are the reheating temperature and the reheating e-fold, respectively. $N_k$ is the number of e-folds after the horizon crossing to the end of the inflationary phase. The constant parameters $\beta$ and $\gamma$ are defined as $\beta = -1/3(1 + w_{re})$ and $\gamma = \beta/(1 - 3 w_{re})$, respectively.

\section{\label{sec:result}
        Consistency with the data}
We study the consistency of the model with the data following two approaches as the exact numerical approach and the standard slow-roll approximations for two coupling functions, introduced in Eq.\eqref{coupling_function}. 

%%%%%%%%%%%%%%%%%%%%%%%%%%%%%%%%%%%%%%%%%%%%%%%%%%%%%%%%%%%%%
%%%%%%%%%%%%%%%%%%%%%%%%%%%%%%%%%%%%%%%%%%%%%%%%%%%%%%%%%%%%%
%%%%%%%%%%%%%%%%%%%%%%%%%%%%%%%%%%%%%%%%%%%%%%%%%%%%%%%%%%%%%
\subsection{Hyperbolic coupling}
\begin{figure}[t]
    \centering
    \subfigure[standard slow-roll]{\includegraphics[width= 6.1cm]{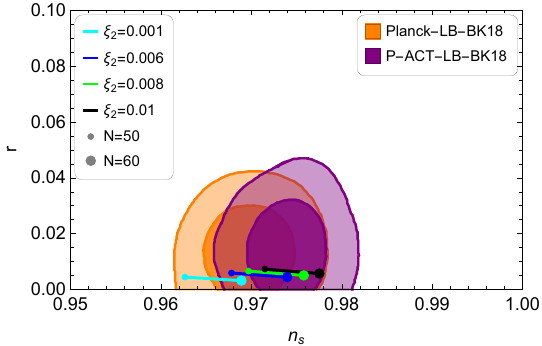}}
    % \subfigure[new slow-roll I]{\includegraphics[width=6.1cm]{figs/rns_s1.pdf}}
    % \subfigure[new slow-roll II]{\includegraphics[width=6.1cm]{figs/rns_s2.pdf}}
    \subfigure[exact numerical \label{rns_tanh_exact}]{\includegraphics[width=6.1cm]{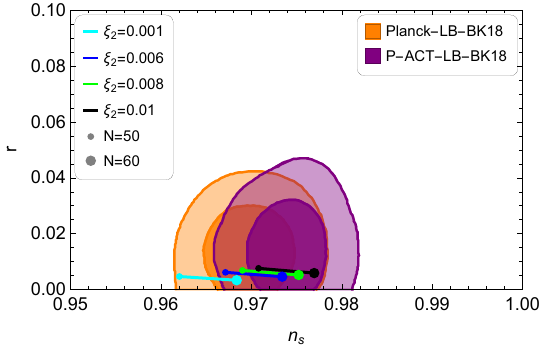}}
    \caption{The $r-n_s$ result for different values of $\xi_2$ is presented, following a) standard slow-roll method, and b) exact numerical method. }
    \label{rns_tanh}
\end{figure}

As the first case, we assumed that the GB coupling is described by a ``tanh" function given in Eq.\eqref{coupling_function}. We start with the standard slow-roll approximations, where it is required to first solve the integral \eqref{N1}. It is assumed that the number of e-folds at the horizon crossing is $N_\star = 0$, and inflation ends at $N=60$. The initial condition is set at the end of inflation, where $\varepsilon$ reaches one, and the field values could be read by solving the relation $\varepsilon_1(\phi_e) = 1$. Solving Eq.\eqref{N1} gives the scalar field as a function of the number of e-folds during inflation. Using this solution, the slow-roll parameters, the scalar spectral index, and the tensor-to-scalar ratio are obtained as a function of the number of e-folds. 
%Repeating the same process for the new slow-roll approximations I and II, one can find the corresponding $n_s$ and $r$. 
Fig.\ref{rns_tanh} exhibits the $r-n_s$ diagram resulting from the slow-roll approach for different values of $\xi_2$, where the bigger/smaller point is related to the result for the number of e-folds $N = 60/50$. It is realized that for $\xi_2 = 0.001$ and $N=60$, the result stays in $1\sigma$ of Planck-LB-BK18; however, it is out of $1\sigma$ of P-ACT-LB-BK18. The result perfectly stands in the $1\sigma$ region of P-ACT-LB-BK18 for the free parameters $\xi_1 = 1$, $\xi_2 = 0.006$, and number of e-fold $N=60$; so that we have $n_s = 0.974$ and $r = 0.0044$. This implies consistency between the model and the ACT data. By considering the figure, it is realized that, for lower values of $\xi_2$, the results lead to the smaller values of $n_s$ and $r$, and if the reduction in $\xi_2$ continues, the result will be out of the $2\sigma$ region of both datasets. The increase in $\xi_2$ leads to a similar conclusion, so that choosing a $\xi_2$ above a specific value results in $n_s$ and $r$ out of observational range.  
%For $\xi_2 = 0.01$, the $r-n_s$ result stands on the edge of the $1\sigma$ boundary of P-ACT-LB-BK18, and higher $\xi_2$ pushes the result out of the $1\sigma$ boundary. Therefore, it could be concluded that the valid range of the free parameter $\xi_2$ is $0.002 < \xi_2 < 0.01$. For this range, the $r-n_s$ result of the Starobinsky potential in EGB gravity stands in the $1\sigma$ boundary of P-ACT-LB-BK18.  
%%%%%%%%%%%%%%%%%%%%%%%%%%%%%%%%%
%%%%%%%%%%%%%%%%%%%%%%%%%%%%%%%%% 
%The results for different slow-roll approximation approaches are almost the same, and there is no essential difference in the values of $r-n_s$. The numerical results about the values of the parameters are presented in the Table.\ref {table}.  \\ 
To get the exact solution, we numerically solve the dynamical equations \eqref{DynSYSN} and determine the background quantities in terms of the number of e-folds. The resulting solutions are substituted in the definition of the slow-roll parameters, Eqs.\eqref{epsilon} and \eqref{delta}, and then, by inserting the slow-roll parameters in the Eq.\eqref{ns_slr}, the scalar spectral index and the tensor-to-scalar ratio are obtained in terms of the number of e-folds. The resulting $r-n_s$ values are displayed in Fig.\ref{rns_tanh_exact} against ACT data for different values of the constant $\xi_2$ and $\xi_1 = 1$. Comparing the result with the slow-roll method, it is realized that for each specific value of $\xi_2$, the obtained $n_s$ in the exact numerical method is a little smaller than that of the slow-roll method. The dependency of the resulting $n_s$ and $r$ on the free parameters $\xi_1$  and $\xi_2$ is the same as the slow-roll methods. As it is shown in the figure, for $\xi_1 = 1$ and $\xi_2 = 0.006$, the results stand in the $1\sigma$ region of the ACT data, and by increasing or decreasing $\xi_2$, the result will gradually move outside of the observational region.
%The allowed range of $\xi_2$ parameter to get consistency with $1\sigma$ is about $<0.0027 < \xi_2 < 0.013$. The free parameter $\xi_2$ also affects the results about the $n_s$ and $r$, so that by increasing the parameter $\xi_1$, one gets higher values for the scalar spectral index and the tensor-to-scalar ratio. For example, for $\xi = 2$, the result stands in $2\sigma$, and for $\xi_1 = 3$, the result would be completely out of the observational range. On the other side, for small values of $\xi_1$, we could get proper consistency with the ACT data by choosing the bigger value of $\xi_2$. 
%%%%%%%%%%%%%%%%%%%%%%%%%%%
\begin{figure}[h]
    \centering
    \includegraphics[width=6cm]{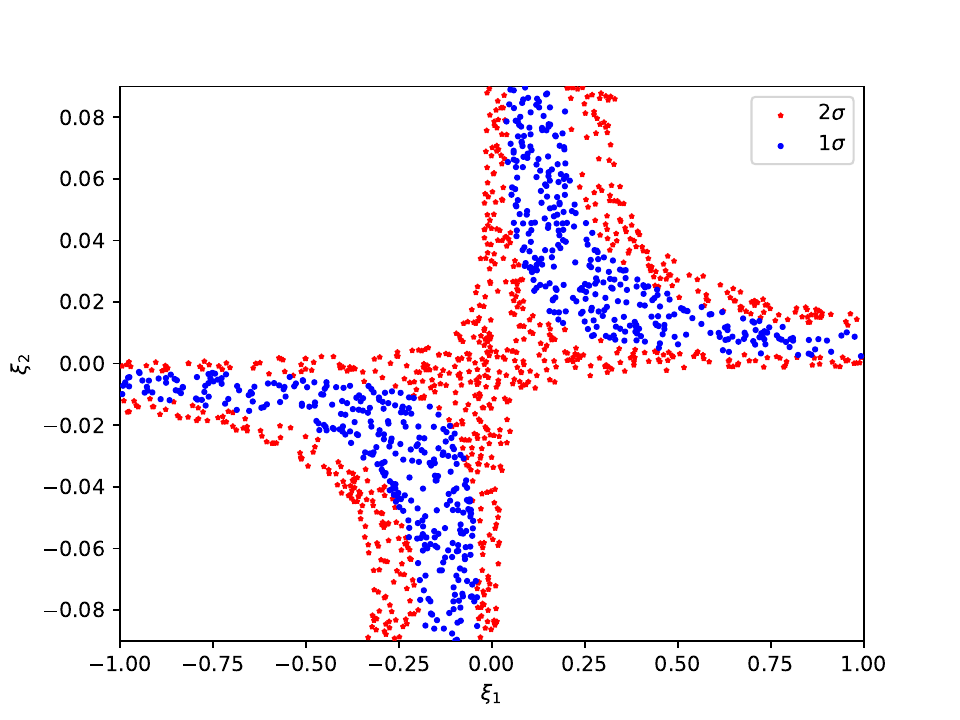}
    \caption{The plot shows the parametric space of $(\xi_1, \xi_2)$ for the case of "tanh" coupling. The blue dots stand for all $(\xi_1, \xi_2)$ points that put the $(n_s, r)$ in the $1\sigma$ region. The combination of blue and red dots shows the set of all $(\xi_1, \xi_2)$ points that put the $(n_s, r)$ in the $2\sigma$ region. It has been assumed that inflation lasts for $60$ e-folds.}
    \label{tanh_parameteric}
\end{figure}
%%%%%%%%%%%%%%%%%%%%%%%%%%%
Fig.\ref{tanh_parameteric} shows the parametric space of the free parameters $\xi_1$ and $\xi_2$. The blue dot shows the set of $(\xi_1. \xi_2)$ points that put the resulting $n_s$ and $r$ in $1\sigma$ region of ACT data, and the combination of blue and red dots are the set of $(\xi_1. \xi_2)$ points that put the result in $2\sigma$ region. The dependency of the result on the free parameters is clear from the figure. It is seen that for $\xi_1 = 1$, the acceptable range of $\xi_2$ is $0.002 < \xi_2 < 0.01$. Note that this range decreases by increasing $\xi_1$, and gets wider for smaller values of $\xi_1$. In addition, one can see a symmetry in the acceptable values of $\xi_1$ and $\xi_2$, so that by $\xi_1, \xi_2 \rightarrow -\xi_1, -\xi_2$, the resulting $n_s$ and $r$ do not change.

The running of the scalar spectral index is another observational data that one can use to check the reliability of an inflationary model by comparing it with. The running is defined as $\alpha_s = dn_s / d\ln k$. For $\xi_1 = 1$, $\xi_2 = 0.006$, and the running is obtained as $\alpha_s = -5\times 10^{-3}$ for the number of e-fold $N = 60$. This value of the running stands in the $1\sigma$ region of the $n_s-\alpha_s$ diagram of ACT; however, it is close to the boundary. \\ 

\begin{figure}
    \centering
    \includegraphics[width=7cm]{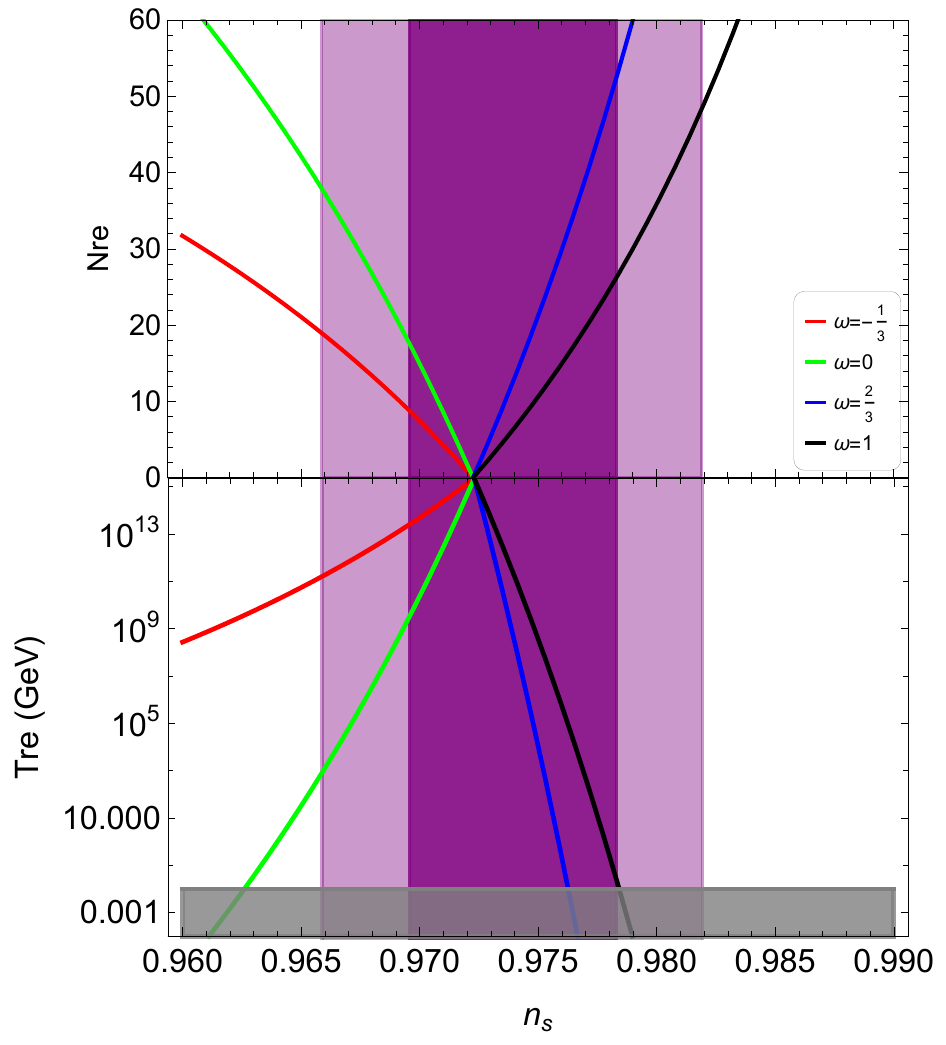}
    \caption{The plot shows the relation of $n_s$ with the reheating phase, so that the top panel shows the behavior of the reheating e-fold $N_{re}$ versus $n_s$, and the bottom panel describes the behavior of the reheating temperature $T_{re}$ versus $n_s$. The result provided for different choice of the reheating equation of state $\omega_{re}$, as $\omega_{re} = -1/3$ (red-curve), $\omega_{re} = 0$ (green-curve), $\omega_{re} = 2/3$ (blue-curve), $\omega_{re} = 1$ (black-curve). The gray area is the excluded region originating from the BBN constraint.}
    \label{tanh_tre}
\end{figure}

After inflation, there is a reheating phase to warm up the universe and provide a smooth transition to the hot Big Bang phase. The influence of the reheating phase on the scalar spectral index, expressed by Eqs.\eqref{Nre} and \eqref{Tre}, is exhibited in Fig.\ref{tanh_tre}, where the free parameters are taken as $\xi_1 = 1$ and $\xi_2 = 0.006$. The top panel indicates the effect of the $N_{re}$ on the scalar spectral index. As it is shown in Fig.\ref{rns_tanh}, for the chosen free parameters, the scalar spectral index is obtained as $n_s \simeq 0.974$ for $ N_k = 60$. This value of $n_s$ corresponds to the reheating e-fold $N_{re} = 14$ and $6$ for $\omega_{re} = 2/3$ and $1$, respectively. The bottom panel displays the behavior of the reheating temperature versus the scalar spectral index. It is realized that, for higher values of reheating equation of state $\omega_{re}$, smaller values of the reheating e-fold $N_{re}$ are achieved, which leads to higher reheating temperature. For instance, for $\omega_{re} = 2/3$ and $1$, the reheating temperature is obtained as $T_{re} \simeq 10^8 \; {\rm GeV}$ and $10^{11} \; {GeV}$. It should be noted that both the ACT data and the BBN temperature ($T_{re} \approx 10 \; {\rm MeV}$) constraints must be satisfied simultaneously. %From the inflationary part, we found that higher values of $\xi_2$ enhance the value of the scalar spectral index. Then, from Fig.\ref{tanh_tre}, this results in higher values for the reheating e-fold and lower reheating temperature, and to satisfy the reheating temperature constraints, bigger values of $\omega_{re}$ are required. 

%%%%%%%%%%%%%%%%%%%%%%%%%%%
%%%%%%%%%%%%%%%%%%%%%%%%%%% 

%%%%%%%%%%%%%%%%%%%%%%%%%%%%%%%%%%%%%%%%%%%%%%%%%%%%%%%%%%%%%
%%%%%%%%%%%%%%%%%%%%%%%%%%%%%%%%%%%%%%%%%%%%%%%%%%%%%%%%%%%%%
%%%%%%%%%%%%%%%%%%%%%%%%%%%%%%%%%%%%%%%%%%%%%%%%%%%%%%%%%%%%%
\subsection{Expotential coupling}
\begin{figure}[t]
    \centering
    \subfigure[standard slow-roll]{\includegraphics[width= 6cm]{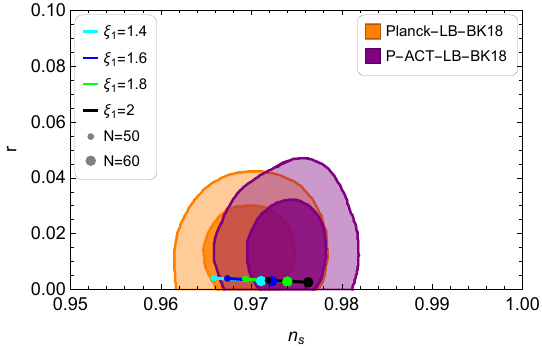}}
    \subfigure[exact numerical \label{sr_rns_exact}]{\includegraphics[width=6cm]{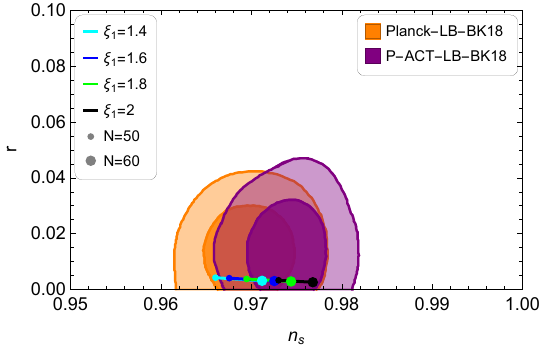}}
    \caption{The plot shows the predicted $r-n_s$ by the model using the exact numerical method and for the exponential coupling. The results are plotted against the ACT $r-n_s$ diagram for different values of the coupling free parameter $\xi_1$. The free parameter $\xi_2$ is taken as $\xi_2 = 1$. Fig a) represents the standard slow roll, and b) denotes the numerical solution obtained by solving the system \eqref{DynSYSN}}
    \label{rns_exp}
\end{figure}

\begin{figure}
    \centering
    \includegraphics[width=6cm]{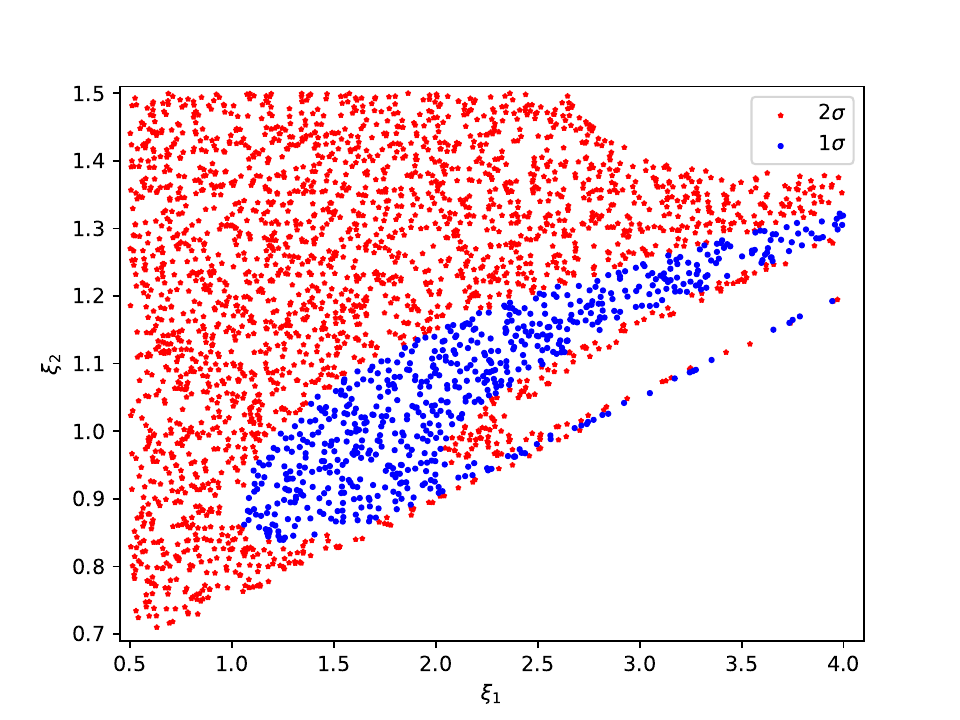}
    \caption{The parametric space of the free parameters $(\xi_1, \xi_2)$ for the case of "exp" coupling is shown, in which the blue dots stand for all $(\xi_1, \xi_2)$ points that put the $(n_s, r)$ in the $1\sigma$ region. The combination of blue and red dots shows the set of all $(\xi_1, \xi_2)$ points that put the $(n_s, r)$ in the $2\sigma$ region. It has been assumed that inflation lasts for $60$ e-folds. }
    \label{exp_parameteric}
\end{figure}

The results for the Starobinsky potential in EGB gravity with an exponential coupling are plotted in Fig.\ref{rns_exp}. The results are presented for different values of the coupling free parameter $\xi_1$. It is realized that, by increasing $\xi_1$, the scalar spectral index increases and the tensor-to-scalar ratio decreases. Then, by increasing $\xi_1$, the result is dragged to the $1\sigma$ region, and if one goes for bigger $\xi_1$, it gets out of the $1\sigma$ region again. The figure illustrates the $r-n_s$ result using the slow-roll and exact numerical methods. The difference between the two approaches is not noticeable, and the resulting values for the scalar spectral index and the tensor-to-scalar ratio are almost the same. The range of $1.3 < \xi_1 < 2.1$ put the results in $1\sigma$ region. 
%%%%%%%%%%%%%%%%%%%%%%%%%%%
%%%%%%%%%%%%%%%%%%%%%%%%%%%
The parametric space of the acceptable pairs of $(\xi_1, \xi_2)$ is plotted in Fig.\ref{exp_parameteric}. The blue dots indicate all $(\xi_1, \xi_2)$ that put the resulting $n_s$ and $r$ in the $1\sigma$ region, and the combination of red and blue points shows all the $(\xi_1, \xi_2)$ which the corresponding $n_s$ and $r$ stay inside the $2\sigma$ region. The acceptable range of $\xi_2$ changes by choosing different values of $\xi_1$, so that for $\xi_1 = 1.8$, the acceptable range of $\xi_2$ parameter is $0.85 < \xi_2 < 1.1$ to put the model in consistency with the $1\sigma$ region of the ACT data. The range gets smaller if we choose a higher value of $\xi_1$, so that for $\xi_1 = 3.8$, one has $1.25 < \xi_2 < 1.3$.

%This result is clear from the parametric space plotted in Fig.\ref{ns_exp_parameteric}, where the parametric space for the free parameters $(\xi_1, \xi_2)$ has been plotted. The figure displays the contour plot of $n_s$ for the range of $0.85 < \xi_1 < 2.7$ and $0.75 < \xi < 1.28$. The points between the lines show the parametric space of $(\xi_1, \xi_2)$ displays values of the free parameter, putting the scalar spectral index in $1\sigma$ of the ACT data. Changing the $xi_2$ affects the acceptable range of $xi_1$, so that for $\xi_2 = 1.15$, the acceptable value of $\xi_1 > 2$, and for $\xi_1 < 0.9$, the valid range of $\xi_1$ gets smaller so that for $\xi_1 < 0.82$, there is no acceptable range of $\xi_1$.  \\ 
%%%%%%%%%%%%%%%%%%%%%%%%%%%
%%%%%%%%%%%%%%%%%%%%%%%%%%%
Considering the running of the scalar spectral index indicates that for $\xi_1 = 1.8$, $\xi_2 = 1$, and the number of e-folds $N = 60$, the running is around $-3.85 \times 10^{-3}$. Compared to the ACT data, this value stands in the $1\sigma$ of the $n_s-\alpha$ diagram. However, it is close to the boundary of the region. \\ 

\begin{figure}[t]
    \centering
    \includegraphics[width=6cm]{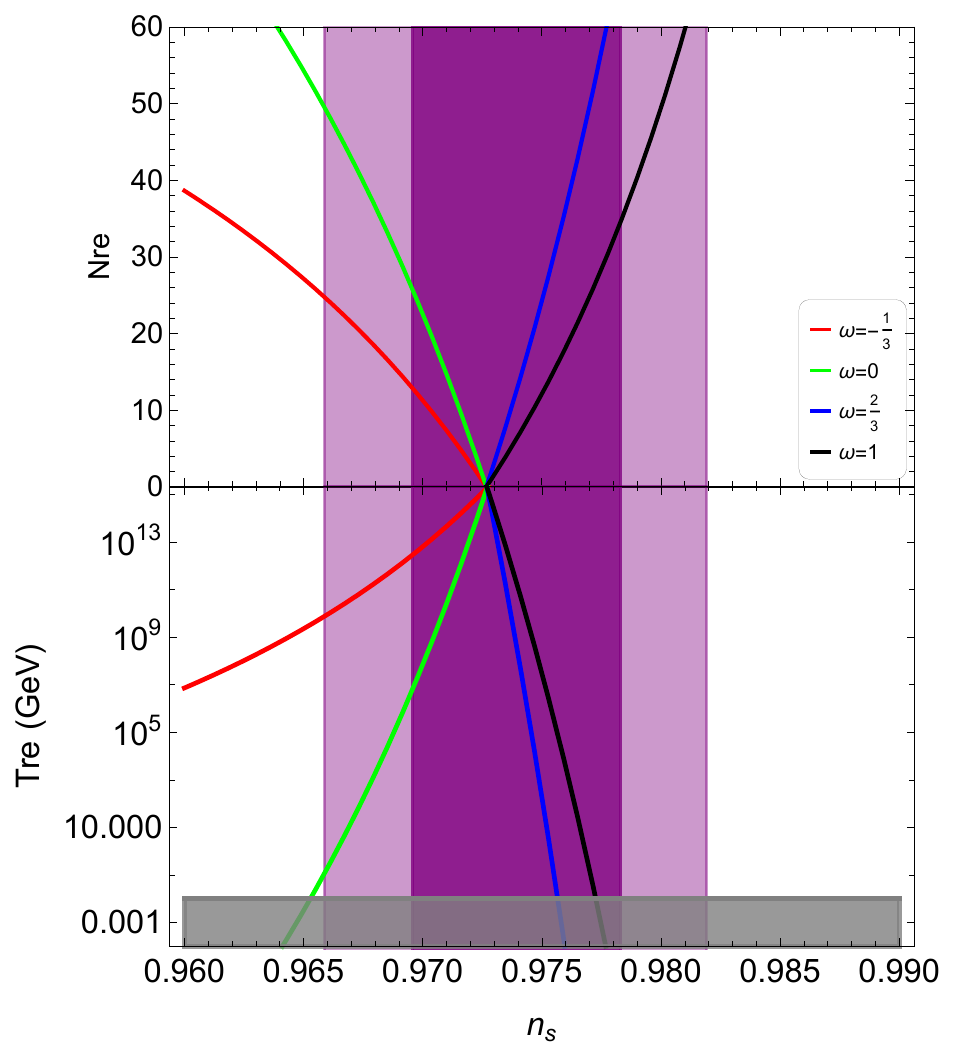}
    \caption{The relation of $n_s$ with the reheating number of e-folds $N_{re}$ is shown by the top panel, and the bottom panel displays the relation between $n_s$ and the reheating temperature $T_{re}$. The plot includes the results for different values of the reheating equation of state parameter as: $\omega_{re} = -1/3$ (red-curve), $\omega_{re} = 0$ (green-curve), $\omega_{re} = 2/3$ (blue-curve), $\omega_{re} = 1$ (black-curve). The gray area is the excluded region originating from the BBN constraint. }
    \label{exp_tre}
\end{figure}

The result of investigating the reheating phase is plotted in Fig.\ref{exp_tre}, where one can find the effect of reheating e-fold and reheating temperature on the scalar spectral index. The plot shows the behavior of the reheating e-fold and reheating temperature versus the scalar spectral index for different values of $\omega_{re}$, and for the free parameters $\xi_1 = 1.8$ and $\xi_1 = 1$. As it is found from Fig.\ref{rns_exp}, for these values of the free parameters, the scalar spectral index is about $n_s = 0.974$. This value of the scalar spectral index corresponds to $N_{re} = 14$ and $6$ respectively for $\omega_{re} = 2/3$ and $1$. Then, increasing $\omega_{re}$ enhances the reheating e-fold, resulting in a lower reheating temperature, as shown in the bottom panel. The reheating temperature is around $T_{re} \simeq 10^8 \; {\rm GeV}$ and $10^{11} \; { \rm GeV}$. 

%%%%%%%%%%%%%%%%%%%%%%%%%%
%%%%%%%%%%%%%%%%%%%%%%%%%%

%%%%%%%%%%%%%%%%%%%%%%%%%%%%%%%%%%%%%%%%%%%%%%%%%%%%%%%%%%%%%
%%%%%%%%%%%%%%%%%%%%%%%%%%%%%%%%%%%%%%%%%%%%%%%%%%%%%%%%%%%%%
%%%%%%%%%%%%%%%%%%%%%%%%%%%%%%%%%%%%%%%%%%%%%%%%%%%%%%%%%%%%%
%%%%%%%%%%%%%%%%%%%%%%%%%%%%%%%%%%%%%%%%%%%%%%%%%%%%%%%%%%%%%
%%%%%%%%%%%%%%%%%%%%%%%%%%%%%%%%%%%%%%%%%%%%%%%%%%%%%%%%%%%%%
%%%%%%%%%%%%%%%%%%%%%%%%%%%%%%%%%%%%%%%%%%%%%%%%%%%%%%%%%%%%%
\section{\label{sec:conclusion}
        Conclusion}
The recent data released by ACT indicates a higher value of the scalar spectral index, i.e. $n_s = 0.9743 \pm 0.0034$, compared to the previous value provided by Planck. This shift in the value of $n_s$ becomes a big challenge for the previously approved inflationary models, so many of them have been disfavored, including the Starobinsky model. 

In this paper, we studied the inflation in the EGB gravity theory by including the Starobinsky potential. The EGB gravity includes quadratic curvature correction to the Einstein-Hilbert action with a non-minimal coupling to the scalar field. We considered two different coupling functions described as $\xi(\phi) \propto \tanh(\xi_2 \phi)$ and $\exp(\xi_2 \phi)$. Due to modified dynamical equations, the model can provide a suitable situation for the potentials disfavored by the constraint given by the ACT data.  \\ 
We followed the slow-roll method and the exact numerical approach to get the model prediction about the scalar spectral index and the tensor-to-scalar ratio. The obtained $r-n_s$ results were plotted against the $r-n_s$ diagram of data for the different values of the coupling free parameters. The resulting $r-n_s$ from the three slow-roll methods are almost the same, and the differences are not essential. The resulting $n_s$ and $r$ out of the model stand in the $1\sigma$ region of the ACT $r-n_s$ diagram, expressing a good consistency between the model and data. The dependency of the $n_s$ and $r$ on the free parameters of the model was considered by exploring the parametric space of $\xi_1$ and $\xi_2$. It was determined that for the case of ``tanh" coupling, the valid range of $\xi_2$ decreases by increasing $\xi_1$, and also, for we observe a symmetry in the result so that by $\xi_1, \xi_2 \rightarrow -\xi_1, -\xi_2$, the resulting $n_s$ and $r$ do not change. The situation for the case of ``exp" coupling is different. There is no such symmetry as ``tanh" case, and also we observe a wider range of $\xi_2$ which keeps the model consistent with the $2\sigma$ region of the data. Considering the running of the scalar spectral index, it was determined that the result of the model stays in the $1\sigma$ region, as well. \\ 
At the final step, we considered the reheating phase and plotted the behavior of reheating e-folds and reheating temperature versus the scalar spectral index. It was determined that to simultaneously satisfy the ACT data $(n_s=0.974)$ and the constraint on the reheating temperature, $\omega_{re}$ should stay in the range $\omega_{re} > 1/3$. Also, it was concluded from the figure that higher values of $\omega_{re}$ correspond to lower reheating e-folds and higher reheating temperature. \\ 
It should also be noted that the model has also been considered using the new slow-roll approximations, proposed in~\cite{Pozdeeva:2024ihc}. The result shows no noticeable difference from the standard slow-roll approximations of the EGB gravity.

%It was determined that for the valid range of the coupling free parameter $0.0012 < \xi_2 \leq 0.01$ the $r-n_s$ values stand in the $1\sigma$ region of the P-ACT-LB-BK18, showing that the Starobinsky potential in the EGB gravity theory can adequately satisfy the constraint provided by the ACT data. 

% The \nocite command causes all entries in a bibliography to be printed out
% whether or not they are actually referenced in the text. This is appropriate
% for the sample file to show the different styles of references, but authors
% most likely will not want to use it.
%\nocite{*}

\bibliography{EGB_ACT}% Produces the bibliography via BibTeX.

%apsrev4-2.bst 2019-01-14 (MD) hand-edited version of apsrev4-1.bst
%Control: key (0)
%Control: author (8) initials jnrlst
%Control: editor formatted (1) identically to author
%Control: production of article title (0) allowed
%Control: page (0) single
%Control: year (1) truncated
%Control: production of eprint (0) enabled
\begin{thebibliography}{128}%
\makeatletter
\providecommand \@ifxundefined [1]{%
 \@ifx{#1\undefined}
}%
\providecommand \@ifnum [1]{%
 \ifnum #1\expandafter \@firstoftwo
 \else \expandafter \@secondoftwo
 \fi
}%
\providecommand \@ifx [1]{%
 \ifx #1\expandafter \@firstoftwo
 \else \expandafter \@secondoftwo
 \fi
}%
\providecommand \natexlab [1]{#1}%
\providecommand \enquote  [1]{``#1''}%
\providecommand \bibnamefont  [1]{#1}%
\providecommand \bibfnamefont [1]{#1}%
\providecommand \citenamefont [1]{#1}%
\providecommand \href@noop [0]{\@secondoftwo}%
\providecommand \href [0]{\begingroup \@sanitize@url \@href}%
\providecommand \@href[1]{\@@startlink{#1}\@@href}%
\providecommand \@@href[1]{\endgroup#1\@@endlink}%
\providecommand \@sanitize@url [0]{\catcode `\\12\catcode `\$12\catcode `\&12\catcode `\#12\catcode `\^12\catcode `\_12\catcode `\%12\relax}%
\providecommand \@@startlink[1]{}%
\providecommand \@@endlink[0]{}%
\providecommand \url  [0]{\begingroup\@sanitize@url \@url }%
\providecommand \@url [1]{\endgroup\@href {#1}{\urlprefix }}%
\providecommand \urlprefix  [0]{URL }%
\providecommand \Eprint [0]{\href }%
\providecommand \doibase [0]{https://doi.org/}%
\providecommand \selectlanguage [0]{\@gobble}%
\providecommand \bibinfo  [0]{\@secondoftwo}%
\providecommand \bibfield  [0]{\@secondoftwo}%
\providecommand \translation [1]{[#1]}%
\providecommand \BibitemOpen [0]{}%
\providecommand \bibitemStop [0]{}%
\providecommand \bibitemNoStop [0]{.\EOS\space}%
\providecommand \EOS [0]{\spacefactor3000\relax}%
\providecommand \BibitemShut  [1]{\csname bibitem#1\endcsname}%
\let\auto@bib@innerbib\@empty
%</preamble>
\bibitem [{\citenamefont {Guth}(1981)}]{Guth:1980zm}%
  \BibitemOpen
  \bibfield  {author} {\bibinfo {author} {\bibfnamefont {A.~H.}\ \bibnamefont {Guth}},\ }\bibfield  {title} {\bibinfo {title} {{The Inflationary Universe: A Possible Solution to the Horizon and Flatness Problems}},\ }\href {https://doi.org/10.1103/PhysRevD.23.347} {\bibfield  {journal} {\bibinfo  {journal} {Phys. Rev. D}\ }\textbf {\bibinfo {volume} {23}},\ \bibinfo {pages} {347} (\bibinfo {year} {1981})}\BibitemShut {NoStop}%
\bibitem [{\citenamefont {Linde}(1982)}]{Linde:1981mu}%
  \BibitemOpen
  \bibfield  {author} {\bibinfo {author} {\bibfnamefont {A.~D.}\ \bibnamefont {Linde}},\ }\bibfield  {title} {\bibinfo {title} {{A New Inflationary Universe Scenario: A Possible Solution of the Horizon, Flatness, Homogeneity, Isotropy and Primordial Monopole Problems}},\ }\href {https://doi.org/10.1016/0370-2693(82)91219-9} {\bibfield  {journal} {\bibinfo  {journal} {Phys. Lett. B}\ }\textbf {\bibinfo {volume} {108}},\ \bibinfo {pages} {389} (\bibinfo {year} {1982})}\BibitemShut {NoStop}%
\bibitem [{\citenamefont {Mukhanov}\ and\ \citenamefont {Chibisov}(1981)}]{Mukhanov:1981xt}%
  \BibitemOpen
  \bibfield  {author} {\bibinfo {author} {\bibfnamefont {V.~F.}\ \bibnamefont {Mukhanov}}\ and\ \bibinfo {author} {\bibfnamefont {G.~V.}\ \bibnamefont {Chibisov}},\ }\bibfield  {title} {\bibinfo {title} {{Quantum Fluctuations and a Nonsingular Universe}},\ }\href@noop {} {\bibfield  {journal} {\bibinfo  {journal} {JETP Lett.}\ }\textbf {\bibinfo {volume} {33}},\ \bibinfo {pages} {532} (\bibinfo {year} {1981})}\BibitemShut {NoStop}%
\bibitem [{\citenamefont {Sato}(1981)}]{Sato:1981qmu}%
  \BibitemOpen
  \bibfield  {author} {\bibinfo {author} {\bibfnamefont {K.}~\bibnamefont {Sato}},\ }\bibfield  {title} {\bibinfo {title} {{First-order phase transition of a vacuum and the expansion of the Universe}},\ }\href {https://doi.org/10.1093/mnras/195.3.467} {\bibfield  {journal} {\bibinfo  {journal} {Mon. Not. Roy. Astron. Soc.}\ }\textbf {\bibinfo {volume} {195}},\ \bibinfo {pages} {467} (\bibinfo {year} {1981})}\BibitemShut {NoStop}%
\bibitem [{\citenamefont {{Starobinsky}}(1996)}]{1996tyli.conf..771S}%
  \BibitemOpen
  \bibfield  {author} {\bibinfo {author} {\bibfnamefont {A.~A.}\ \bibnamefont {{Starobinsky}}},\ }\bibfield  {title} {\bibinfo {title} {{A New Type of Isotropic Cosmological Models without Singularity - Phys. Lett. B91, 99 (1980)}},\ }in\ \href@noop {} {\emph {\bibinfo {booktitle} {30 Years of the Landau Institute}}},\ Vol.~\bibinfo {volume} {11},\ \bibinfo {editor} {edited by\ \bibinfo {editor} {\bibfnamefont {I.~M.}\ \bibnamefont {{Khalatnikov}}}\ and\ \bibinfo {editor} {\bibfnamefont {V.~P.}\ \bibnamefont {{Mineev}}}}\ (\bibinfo {year} {1996})\ p.\ \bibinfo {pages} {771}\BibitemShut {NoStop}%
\bibitem [{\citenamefont {Albrecht}\ and\ \citenamefont {Steinhardt}(1982)}]{PhysRevLett.48.1220}%
  \BibitemOpen
  \bibfield  {author} {\bibinfo {author} {\bibfnamefont {A.}~\bibnamefont {Albrecht}}\ and\ \bibinfo {author} {\bibfnamefont {P.~J.}\ \bibnamefont {Steinhardt}},\ }\bibfield  {title} {\bibinfo {title} {Cosmology for grand unified theories with radiatively induced symmetry breaking},\ }\href {https://doi.org/10.1103/PhysRevLett.48.1220} {\bibfield  {journal} {\bibinfo  {journal} {Phys. Rev. Lett.}\ }\textbf {\bibinfo {volume} {48}},\ \bibinfo {pages} {1220} (\bibinfo {year} {1982})}\BibitemShut {NoStop}%
\bibitem [{\citenamefont {Starobinsky}(1982)}]{Starobinsky:1982ee}%
  \BibitemOpen
  \bibfield  {author} {\bibinfo {author} {\bibfnamefont {A.~A.}\ \bibnamefont {Starobinsky}},\ }\bibfield  {title} {\bibinfo {title} {{Dynamics of Phase Transition in the New Inflationary Universe Scenario and Generation of Perturbations}},\ }\href {https://doi.org/10.1016/0370-2693(82)90541-X} {\bibfield  {journal} {\bibinfo  {journal} {Phys. Lett. B}\ }\textbf {\bibinfo {volume} {117}},\ \bibinfo {pages} {175} (\bibinfo {year} {1982})}\BibitemShut {NoStop}%
\bibitem [{\citenamefont {Hinshaw}\ \emph {et~al.}(2013)\citenamefont {Hinshaw} \emph {et~al.}}]{WMAP:2012nax}%
  \BibitemOpen
  \bibfield  {author} {\bibinfo {author} {\bibfnamefont {G.}~\bibnamefont {Hinshaw}} \emph {et~al.} (\bibinfo {collaboration} {WMAP}),\ }\bibfield  {title} {\bibinfo {title} {{Nine-Year Wilkinson Microwave Anisotropy Probe (WMAP) Observations: Cosmological Parameter Results}},\ }\href {https://doi.org/10.1088/0067-0049/208/2/19} {\bibfield  {journal} {\bibinfo  {journal} {Astrophys. J. Suppl.}\ }\textbf {\bibinfo {volume} {208}},\ \bibinfo {pages} {19} (\bibinfo {year} {2013})},\ \Eprint {https://arxiv.org/abs/1212.5226} {arXiv:1212.5226 [astro-ph.CO]} \BibitemShut {NoStop}%
\bibitem [{\citenamefont {Ade}\ \emph {et~al.}(2016)\citenamefont {Ade} \emph {et~al.}}]{Planck:2015sxf}%
  \BibitemOpen
  \bibfield  {author} {\bibinfo {author} {\bibfnamefont {P.~A.~R.}\ \bibnamefont {Ade}} \emph {et~al.} (\bibinfo {collaboration} {Planck}),\ }\bibfield  {title} {\bibinfo {title} {{Planck 2015 results. XX. Constraints on inflation}},\ }\href {https://doi.org/10.1051/0004-6361/201525898} {\bibfield  {journal} {\bibinfo  {journal} {Astron. Astrophys.}\ }\textbf {\bibinfo {volume} {594}},\ \bibinfo {pages} {A20} (\bibinfo {year} {2016})},\ \Eprint {https://arxiv.org/abs/1502.02114} {arXiv:1502.02114 [astro-ph.CO]} \BibitemShut {NoStop}%
\bibitem [{\citenamefont {Akrami}\ \emph {et~al.}(2020)\citenamefont {Akrami} \emph {et~al.}}]{Planck:2018jri}%
  \BibitemOpen
  \bibfield  {author} {\bibinfo {author} {\bibfnamefont {Y.}~\bibnamefont {Akrami}} \emph {et~al.} (\bibinfo {collaboration} {Planck}),\ }\bibfield  {title} {\bibinfo {title} {{Planck 2018 results. X. Constraints on inflation}},\ }\href {https://doi.org/10.1051/0004-6361/201833887} {\bibfield  {journal} {\bibinfo  {journal} {Astron. Astrophys.}\ }\textbf {\bibinfo {volume} {641}},\ \bibinfo {pages} {A10} (\bibinfo {year} {2020})},\ \Eprint {https://arxiv.org/abs/1807.06211} {arXiv:1807.06211 [astro-ph.CO]} \BibitemShut {NoStop}%
\bibitem [{\citenamefont {Barenboim}\ and\ \citenamefont {Kinney}(2007)}]{Barenboim:2007ii}%
  \BibitemOpen
  \bibfield  {author} {\bibinfo {author} {\bibfnamefont {G.}~\bibnamefont {Barenboim}}\ and\ \bibinfo {author} {\bibfnamefont {W.~H.}\ \bibnamefont {Kinney}},\ }\bibfield  {title} {\bibinfo {title} {{Slow roll in simple non-canonical inflation}},\ }\href {https://doi.org/10.1088/1475-7516/2007/03/014} {\bibfield  {journal} {\bibinfo  {journal} {JCAP}\ }\textbf {\bibinfo {volume} {0703}},\ \bibinfo {pages} {014}},\ \Eprint {https://arxiv.org/abs/astro-ph/0701343} {arXiv:astro-ph/0701343 [astro-ph]} \BibitemShut {NoStop}%
%%CITATION = ASTRO-PH/0701343;%%
\bibitem [{\citenamefont {Franche}\ \emph {et~al.}(2010)\citenamefont {Franche}, \citenamefont {Gwyn}, \citenamefont {Underwood},\ and\ \citenamefont {Wissanji}}]{Franche:2010yj}%
  \BibitemOpen
  \bibfield  {author} {\bibinfo {author} {\bibfnamefont {P.}~\bibnamefont {Franche}}, \bibinfo {author} {\bibfnamefont {R.}~\bibnamefont {Gwyn}}, \bibinfo {author} {\bibfnamefont {B.}~\bibnamefont {Underwood}},\ and\ \bibinfo {author} {\bibfnamefont {A.}~\bibnamefont {Wissanji}},\ }\bibfield  {title} {\bibinfo {title} {{Initial Conditions for Non-Canonical Inflation}},\ }\href {https://doi.org/10.1103/PhysRevD.82.063528} {\bibfield  {journal} {\bibinfo  {journal} {Phys. Rev.}\ }\textbf {\bibinfo {volume} {D82}},\ \bibinfo {pages} {063528} (\bibinfo {year} {2010})},\ \Eprint {https://arxiv.org/abs/1002.2639} {arXiv:1002.2639 [hep-th]} \BibitemShut {NoStop}%
%%CITATION = ARXIV:1002.2639;%%
\bibitem [{\citenamefont {Fairbairn}\ and\ \citenamefont {Tytgat}(2002)}]{Fairbairn:2002yp}%
  \BibitemOpen
  \bibfield  {author} {\bibinfo {author} {\bibfnamefont {M.}~\bibnamefont {Fairbairn}}\ and\ \bibinfo {author} {\bibfnamefont {M.~H.~G.}\ \bibnamefont {Tytgat}},\ }\bibfield  {title} {\bibinfo {title} {{Inflation from a tachyon fluid?}},\ }\href {https://doi.org/10.1016/S0370-2693(02)02638-2} {\bibfield  {journal} {\bibinfo  {journal} {Phys. Lett.}\ }\textbf {\bibinfo {volume} {B546}},\ \bibinfo {pages} {1} (\bibinfo {year} {2002})},\ \Eprint {https://arxiv.org/abs/hep-th/0204070} {arXiv:hep-th/0204070 [hep-th]} \BibitemShut {NoStop}%
%%CITATION = HEP-TH/0204070;%%
\bibitem [{\citenamefont {Mukohyama}(2002)}]{Mukohyama:2002cn}%
  \BibitemOpen
  \bibfield  {author} {\bibinfo {author} {\bibfnamefont {S.}~\bibnamefont {Mukohyama}},\ }\bibfield  {title} {\bibinfo {title} {{Brane cosmology driven by the rolling tachyon}},\ }\href {https://doi.org/10.1103/PhysRevD.66.024009} {\bibfield  {journal} {\bibinfo  {journal} {Phys. Rev.}\ }\textbf {\bibinfo {volume} {D66}},\ \bibinfo {pages} {024009} (\bibinfo {year} {2002})},\ \Eprint {https://arxiv.org/abs/hep-th/0204084} {arXiv:hep-th/0204084 [hep-th]} \BibitemShut {NoStop}%
%%CITATION = HEP-TH/0204084;%%
\bibitem [{\citenamefont {Feinstein}(2002)}]{Feinstein:2002aj}%
  \BibitemOpen
  \bibfield  {author} {\bibinfo {author} {\bibfnamefont {A.}~\bibnamefont {Feinstein}},\ }\bibfield  {title} {\bibinfo {title} {{Power law inflation from the rolling tachyon}},\ }\href {https://doi.org/10.1103/PhysRevD.66.063511} {\bibfield  {journal} {\bibinfo  {journal} {Phys. Rev.}\ }\textbf {\bibinfo {volume} {D66}},\ \bibinfo {pages} {063511} (\bibinfo {year} {2002})},\ \Eprint {https://arxiv.org/abs/hep-th/0204140} {arXiv:hep-th/0204140 [hep-th]} \BibitemShut {NoStop}%
%%CITATION = HEP-TH/0204140;%%
\bibitem [{\citenamefont {Padmanabhan}(2002)}]{Padmanabhan:2002cp}%
  \BibitemOpen
  \bibfield  {author} {\bibinfo {author} {\bibfnamefont {T.}~\bibnamefont {Padmanabhan}},\ }\bibfield  {title} {\bibinfo {title} {{Accelerated expansion of the universe driven by tachyonic matter}},\ }\href {https://doi.org/10.1103/PhysRevD.66.021301} {\bibfield  {journal} {\bibinfo  {journal} {Phys. Rev.}\ }\textbf {\bibinfo {volume} {D66}},\ \bibinfo {pages} {021301} (\bibinfo {year} {2002})},\ \Eprint {https://arxiv.org/abs/hep-th/0204150} {arXiv:hep-th/0204150 [hep-th]} \BibitemShut {NoStop}%
%%CITATION = HEP-TH/0204150;%%
\bibitem [{\citenamefont {Spalinski}(2007)}]{Spalinski:2007dv}%
  \BibitemOpen
  \bibfield  {author} {\bibinfo {author} {\bibfnamefont {M.}~\bibnamefont {Spalinski}},\ }\bibfield  {title} {\bibinfo {title} {{On Power law inflation in DBI models}},\ }\href {https://doi.org/10.1088/1475-7516/2007/05/017} {\bibfield  {journal} {\bibinfo  {journal} {JCAP}\ }\textbf {\bibinfo {volume} {0705}},\ \bibinfo {pages} {017}},\ \Eprint {https://arxiv.org/abs/hep-th/0702196} {arXiv:hep-th/0702196 [hep-th]} \BibitemShut {NoStop}%
%%CITATION = HEP-TH/0702196;%%
\bibitem [{\citenamefont {Bessada}\ \emph {et~al.}(2009)\citenamefont {Bessada}, \citenamefont {Kinney},\ and\ \citenamefont {Tzirakis}}]{Bessada:2009pe}%
  \BibitemOpen
  \bibfield  {author} {\bibinfo {author} {\bibfnamefont {D.}~\bibnamefont {Bessada}}, \bibinfo {author} {\bibfnamefont {W.~H.}\ \bibnamefont {Kinney}},\ and\ \bibinfo {author} {\bibfnamefont {K.}~\bibnamefont {Tzirakis}},\ }\bibfield  {title} {\bibinfo {title} {{Inflationary potentials in DBI models}},\ }\href {https://doi.org/10.1088/1475-7516/2009/09/031} {\bibfield  {journal} {\bibinfo  {journal} {JCAP}\ }\textbf {\bibinfo {volume} {0909}},\ \bibinfo {pages} {031}},\ \Eprint {https://arxiv.org/abs/0907.1311} {arXiv:0907.1311 [gr-qc]} \BibitemShut {NoStop}%
%%CITATION = ARXIV:0907.1311;%%
\bibitem [{\citenamefont {Weller}\ \emph {et~al.}(2012)\citenamefont {Weller}, \citenamefont {van~de Bruck},\ and\ \citenamefont {Mota}}]{Weller:2011ey}%
  \BibitemOpen
  \bibfield  {author} {\bibinfo {author} {\bibfnamefont {J.~M.}\ \bibnamefont {Weller}}, \bibinfo {author} {\bibfnamefont {C.}~\bibnamefont {van~de Bruck}},\ and\ \bibinfo {author} {\bibfnamefont {D.~F.}\ \bibnamefont {Mota}},\ }\bibfield  {title} {\bibinfo {title} {{Inflationary predictions in scalar-tensor DBI inflation}},\ }\href {https://doi.org/10.1088/1475-7516/2012/06/002} {\bibfield  {journal} {\bibinfo  {journal} {JCAP}\ }\textbf {\bibinfo {volume} {1206}},\ \bibinfo {pages} {002}},\ \Eprint {https://arxiv.org/abs/1111.0237} {arXiv:1111.0237 [astro-ph.CO]} \BibitemShut {NoStop}%
%%CITATION = ARXIV:1111.0237;%%
\bibitem [{\citenamefont {Maeda}\ and\ \citenamefont {Yamamoto}(2013)}]{maeda2013stability}%
  \BibitemOpen
  \bibfield  {author} {\bibinfo {author} {\bibfnamefont {K.-i.}\ \bibnamefont {Maeda}}\ and\ \bibinfo {author} {\bibfnamefont {K.}~\bibnamefont {Yamamoto}},\ }\bibfield  {title} {\bibinfo {title} {Stability analysis of inflation with an su (2) gauge field},\ }\href@noop {} {\bibfield  {journal} {\bibinfo  {journal} {Journal of Cosmology and Astroparticle Physics}\ }\textbf {\bibinfo {volume} {2013}}\bibinfo  {number} { (12)},\ \bibinfo {pages} {018}}\BibitemShut {NoStop}%
\bibitem [{\citenamefont {Alexander}\ \emph {et~al.}(2015)\citenamefont {Alexander}, \citenamefont {Jyoti}, \citenamefont {Kosowsky},\ and\ \citenamefont {Marcian{\`o}}}]{alexander2015dynamics}%
  \BibitemOpen
\bibfield  {number} {  }\bibfield  {author} {\bibinfo {author} {\bibfnamefont {S.}~\bibnamefont {Alexander}}, \bibinfo {author} {\bibfnamefont {D.}~\bibnamefont {Jyoti}}, \bibinfo {author} {\bibfnamefont {A.}~\bibnamefont {Kosowsky}},\ and\ \bibinfo {author} {\bibfnamefont {A.}~\bibnamefont {Marcian{\`o}}},\ }\bibfield  {title} {\bibinfo {title} {Dynamics of gauge field inflation},\ }\href@noop {} {\bibfield  {journal} {\bibinfo  {journal} {Journal of Cosmology and Astroparticle Physics}\ }\textbf {\bibinfo {volume} {2015}}\bibinfo  {number} { (05)},\ \bibinfo {pages} {005}}\BibitemShut {NoStop}%
\bibitem [{\citenamefont {Maartens}\ \emph {et~al.}(2000)\citenamefont {Maartens}, \citenamefont {Wands}, \citenamefont {Bassett},\ and\ \citenamefont {Heard}}]{maartens2000chaotic}%
  \BibitemOpen
\bibfield  {number} {  }\bibfield  {author} {\bibinfo {author} {\bibfnamefont {R.}~\bibnamefont {Maartens}}, \bibinfo {author} {\bibfnamefont {D.}~\bibnamefont {Wands}}, \bibinfo {author} {\bibfnamefont {B.~A.}\ \bibnamefont {Bassett}},\ and\ \bibinfo {author} {\bibfnamefont {I.~P.}\ \bibnamefont {Heard}},\ }\bibfield  {title} {\bibinfo {title} {Chaotic inflation on the brane},\ }\href@noop {} {\bibfield  {journal} {\bibinfo  {journal} {Physical Review D}\ }\textbf {\bibinfo {volume} {62}},\ \bibinfo {pages} {041301} (\bibinfo {year} {2000})}\BibitemShut {NoStop}%
\bibitem [{\citenamefont {Mohammadi}\ \emph {et~al.}(2022{\natexlab{a}})\citenamefont {Mohammadi}, \citenamefont {Golanbari}, \citenamefont {Nasri},\ and\ \citenamefont {Saaidi}}]{Mohammadi:2020ake}%
  \BibitemOpen
  \bibfield  {author} {\bibinfo {author} {\bibfnamefont {A.}~\bibnamefont {Mohammadi}}, \bibinfo {author} {\bibfnamefont {T.}~\bibnamefont {Golanbari}}, \bibinfo {author} {\bibfnamefont {S.}~\bibnamefont {Nasri}},\ and\ \bibinfo {author} {\bibfnamefont {K.}~\bibnamefont {Saaidi}},\ }\bibfield  {title} {\bibinfo {title} {{Brane inflation: Swampland criteria, TCC, and reheating predictions}},\ }\href {https://doi.org/10.1016/j.astropartphys.2022.102734} {\bibfield  {journal} {\bibinfo  {journal} {Astropart. Phys.}\ }\textbf {\bibinfo {volume} {142}},\ \bibinfo {pages} {102734} (\bibinfo {year} {2022}{\natexlab{a}})},\ \Eprint {https://arxiv.org/abs/2006.09489} {arXiv:2006.09489 [gr-qc]} \BibitemShut {NoStop}%
\bibitem [{\citenamefont {Mohammadi}\ \emph {et~al.}(2021{\natexlab{a}})\citenamefont {Mohammadi}, \citenamefont {Golanbari},\ and\ \citenamefont {Enayati}}]{Mohammadi:2020ctd}%
  \BibitemOpen
  \bibfield  {author} {\bibinfo {author} {\bibfnamefont {A.}~\bibnamefont {Mohammadi}}, \bibinfo {author} {\bibfnamefont {T.}~\bibnamefont {Golanbari}},\ and\ \bibinfo {author} {\bibfnamefont {J.}~\bibnamefont {Enayati}},\ }\bibfield  {title} {\bibinfo {title} {{Brane inflation and trans-Planckian censorship conjecture}},\ }\href {https://doi.org/10.1103/PhysRevD.104.123515} {\bibfield  {journal} {\bibinfo  {journal} {Phys. Rev. D}\ }\textbf {\bibinfo {volume} {104}},\ \bibinfo {pages} {123515} (\bibinfo {year} {2021}{\natexlab{a}})},\ \Eprint {https://arxiv.org/abs/2012.01512} {arXiv:2012.01512 [hep-th]} \BibitemShut {NoStop}%
\bibitem [{\citenamefont {Berera}(1995)}]{berera1995warm}%
  \BibitemOpen
  \bibfield  {author} {\bibinfo {author} {\bibfnamefont {A.}~\bibnamefont {Berera}},\ }\bibfield  {title} {\bibinfo {title} {Warm inflation},\ }\href@noop {} {\bibfield  {journal} {\bibinfo  {journal} {Physical Review Letters}\ }\textbf {\bibinfo {volume} {75}},\ \bibinfo {pages} {3218} (\bibinfo {year} {1995})}\BibitemShut {NoStop}%
\bibitem [{\citenamefont {Sheikhahmadi}\ \emph {et~al.}(2019)\citenamefont {Sheikhahmadi}, \citenamefont {Mohammadi}, \citenamefont {Aghamohammadi}, \citenamefont {Harko}, \citenamefont {Herrera}, \citenamefont {Corda}, \citenamefont {Abebe},\ and\ \citenamefont {Saaidi}}]{Sheikhahmadi:2019gzs}%
  \BibitemOpen
  \bibfield  {author} {\bibinfo {author} {\bibfnamefont {H.}~\bibnamefont {Sheikhahmadi}}, \bibinfo {author} {\bibfnamefont {A.}~\bibnamefont {Mohammadi}}, \bibinfo {author} {\bibfnamefont {A.}~\bibnamefont {Aghamohammadi}}, \bibinfo {author} {\bibfnamefont {T.}~\bibnamefont {Harko}}, \bibinfo {author} {\bibfnamefont {R.}~\bibnamefont {Herrera}}, \bibinfo {author} {\bibfnamefont {C.}~\bibnamefont {Corda}}, \bibinfo {author} {\bibfnamefont {A.}~\bibnamefont {Abebe}},\ and\ \bibinfo {author} {\bibfnamefont {K.}~\bibnamefont {Saaidi}},\ }\bibfield  {title} {\bibinfo {title} {{Constraining chameleon field driven warm inflation with Planck 2018 data}},\ }\href {https://doi.org/10.1140/epjc/s10052-019-7571-0} {\bibfield  {journal} {\bibinfo  {journal} {Eur. Phys. J.}\ }\textbf {\bibinfo {volume} {C79}},\ \bibinfo {pages} {1038} (\bibinfo {year} {2019})},\ \Eprint {https://arxiv.org/abs/1907.10966} {arXiv:1907.10966 [gr-qc]} \BibitemShut {NoStop}%
%%CITATION = ARXIV:1907.10966;%%
\bibitem [{\citenamefont {Mohammadi}\ \emph {et~al.}(2018)\citenamefont {Mohammadi}, \citenamefont {Saaidi},\ and\ \citenamefont {Golanbari}}]{Mohammadi:2018oku}%
  \BibitemOpen
  \bibfield  {author} {\bibinfo {author} {\bibfnamefont {A.}~\bibnamefont {Mohammadi}}, \bibinfo {author} {\bibfnamefont {K.}~\bibnamefont {Saaidi}},\ and\ \bibinfo {author} {\bibfnamefont {T.}~\bibnamefont {Golanbari}},\ }\bibfield  {title} {\bibinfo {title} {{Tachyon constant-roll inflation}},\ }\href {https://doi.org/10.1103/PhysRevD.97.083006} {\bibfield  {journal} {\bibinfo  {journal} {Phys. Rev.}\ }\textbf {\bibinfo {volume} {D97}},\ \bibinfo {pages} {083006} (\bibinfo {year} {2018})},\ \Eprint {https://arxiv.org/abs/1801.03487} {arXiv:1801.03487 [hep-ph]} \BibitemShut {NoStop}%
%%CITATION = ARXIV:1801.03487;%%
\bibitem [{\citenamefont {Mohammadi}\ \emph {et~al.}(2019)\citenamefont {Mohammadi}, \citenamefont {Saaidi},\ and\ \citenamefont {Sheikhahmadi}}]{Mohammadi:2019dpu}%
  \BibitemOpen
  \bibfield  {author} {\bibinfo {author} {\bibfnamefont {A.}~\bibnamefont {Mohammadi}}, \bibinfo {author} {\bibfnamefont {K.}~\bibnamefont {Saaidi}},\ and\ \bibinfo {author} {\bibfnamefont {H.}~\bibnamefont {Sheikhahmadi}},\ }\bibfield  {title} {\bibinfo {title} {{Constant-roll approach to non-canonical inflation}},\ }\href {https://doi.org/10.1103/PhysRevD.100.083520} {\bibfield  {journal} {\bibinfo  {journal} {Phys. Rev.}\ }\textbf {\bibinfo {volume} {D100}},\ \bibinfo {pages} {083520} (\bibinfo {year} {2019})},\ \Eprint {https://arxiv.org/abs/1803.01715} {arXiv:1803.01715 [astro-ph.CO]} \BibitemShut {NoStop}%
%%CITATION = ARXIV:1803.01715;%%
\bibitem [{\citenamefont {Golanbari}\ \emph {et~al.}(2020)\citenamefont {Golanbari}, \citenamefont {Mohammadi},\ and\ \citenamefont {Saaidi}}]{Mohammadi:2018zkf}%
  \BibitemOpen
  \bibfield  {author} {\bibinfo {author} {\bibfnamefont {T.}~\bibnamefont {Golanbari}}, \bibinfo {author} {\bibfnamefont {A.}~\bibnamefont {Mohammadi}},\ and\ \bibinfo {author} {\bibfnamefont {K.}~\bibnamefont {Saaidi}},\ }\bibfield  {title} {\bibinfo {title} {{Observational constraints on DBI constant-roll inflation}},\ }\href {https://doi.org/10.1016/j.dark.2019.100456} {\bibfield  {journal} {\bibinfo  {journal} {Phys. Dark Univ.}\ }\textbf {\bibinfo {volume} {27}},\ \bibinfo {pages} {100456} (\bibinfo {year} {2020})},\ \Eprint {https://arxiv.org/abs/1808.07246} {arXiv:1808.07246 [gr-qc]} \BibitemShut {NoStop}%
%%CITATION = ARXIV:1808.07246;%%
\bibitem [{\citenamefont {Mohammadi}\ \emph {et~al.}(2020{\natexlab{a}})\citenamefont {Mohammadi}, \citenamefont {Golanbari},\ and\ \citenamefont {Saaidi}}]{Mohammadi:2019qeu}%
  \BibitemOpen
  \bibfield  {author} {\bibinfo {author} {\bibfnamefont {A.}~\bibnamefont {Mohammadi}}, \bibinfo {author} {\bibfnamefont {T.}~\bibnamefont {Golanbari}},\ and\ \bibinfo {author} {\bibfnamefont {K.}~\bibnamefont {Saaidi}},\ }\bibfield  {title} {\bibinfo {title} {{Beta-function formalism for k-essence constant-roll inflation}},\ }\href {https://doi.org/10.1016/j.dark.2020.100505} {\bibfield  {journal} {\bibinfo  {journal} {Phys. Dark Univ.}\ }\textbf {\bibinfo {volume} {28}},\ \bibinfo {pages} {100505} (\bibinfo {year} {2020}{\natexlab{a}})},\ \Eprint {https://arxiv.org/abs/1912.07006} {arXiv:1912.07006 [gr-qc]} \BibitemShut {NoStop}%
%%CITATION = ARXIV:1912.07006;%%
\bibitem [{\citenamefont {Mohammadi}\ \emph {et~al.}(2020{\natexlab{b}})\citenamefont {Mohammadi}, \citenamefont {Golanbari}, \citenamefont {Nasri},\ and\ \citenamefont {Saaidi}}]{Mohammadi:2020ftb}%
  \BibitemOpen
  \bibfield  {author} {\bibinfo {author} {\bibfnamefont {A.}~\bibnamefont {Mohammadi}}, \bibinfo {author} {\bibfnamefont {T.}~\bibnamefont {Golanbari}}, \bibinfo {author} {\bibfnamefont {S.}~\bibnamefont {Nasri}},\ and\ \bibinfo {author} {\bibfnamefont {K.}~\bibnamefont {Saaidi}},\ }\bibfield  {title} {\bibinfo {title} {{Constant-roll brane inflation}},\ }\href {https://doi.org/10.1103/PhysRevD.101.123537} {\bibfield  {journal} {\bibinfo  {journal} {Phys. Rev. D}\ }\textbf {\bibinfo {volume} {101}},\ \bibinfo {pages} {123537} (\bibinfo {year} {2020}{\natexlab{b}})},\ \Eprint {https://arxiv.org/abs/2004.12137} {arXiv:2004.12137 [gr-qc]} \BibitemShut {NoStop}%
\bibitem [{\citenamefont {Mohammadi}\ \emph {et~al.}(2021{\natexlab{b}})\citenamefont {Mohammadi}, \citenamefont {Golanbari}, \citenamefont {Bamba},\ and\ \citenamefont {Lobo}}]{Mohammadi:2021wde}%
  \BibitemOpen
  \bibfield  {author} {\bibinfo {author} {\bibfnamefont {A.}~\bibnamefont {Mohammadi}}, \bibinfo {author} {\bibfnamefont {T.}~\bibnamefont {Golanbari}}, \bibinfo {author} {\bibfnamefont {K.}~\bibnamefont {Bamba}},\ and\ \bibinfo {author} {\bibfnamefont {I.~P.}\ \bibnamefont {Lobo}},\ }\bibfield  {title} {\bibinfo {title} {{Tsallis holographic dark energy for inflation}},\ }\href {https://doi.org/10.1103/PhysRevD.103.083505} {\bibfield  {journal} {\bibinfo  {journal} {Phys. Rev. D}\ }\textbf {\bibinfo {volume} {103}},\ \bibinfo {pages} {083505} (\bibinfo {year} {2021}{\natexlab{b}})},\ \Eprint {https://arxiv.org/abs/2101.06378} {arXiv:2101.06378 [gr-qc]} \BibitemShut {NoStop}%
\bibitem [{\citenamefont {Mohammadi}(2021)}]{Mohammadi:2021gvf}%
  \BibitemOpen
  \bibfield  {author} {\bibinfo {author} {\bibfnamefont {A.}~\bibnamefont {Mohammadi}},\ }\bibfield  {title} {\bibinfo {title} {{Holographic warm inflation}},\ }\href {https://doi.org/10.1103/PhysRevD.104.123538} {\bibfield  {journal} {\bibinfo  {journal} {Phys. Rev. D}\ }\textbf {\bibinfo {volume} {104}},\ \bibinfo {pages} {123538} (\bibinfo {year} {2021})},\ \Eprint {https://arxiv.org/abs/2109.00247} {arXiv:2109.00247 [gr-qc]} \BibitemShut {NoStop}%
\bibitem [{\citenamefont {Mohammadi}(2022)}]{Mohammadi:2022vru}%
  \BibitemOpen
  \bibfield  {author} {\bibinfo {author} {\bibfnamefont {A.}~\bibnamefont {Mohammadi}},\ }\bibfield  {title} {\bibinfo {title} {{Constant-roll inflation driven by holographic dark energy}},\ }\href {https://doi.org/10.1016/j.dark.2022.101055} {\bibfield  {journal} {\bibinfo  {journal} {Phys. Dark Univ.}\ }\textbf {\bibinfo {volume} {36}},\ \bibinfo {pages} {101055} (\bibinfo {year} {2022})},\ \Eprint {https://arxiv.org/abs/2203.06643} {arXiv:2203.06643 [gr-qc]} \BibitemShut {NoStop}%
\bibitem [{\citenamefont {Mohammadi}\ \emph {et~al.}(2022{\natexlab{b}})\citenamefont {Mohammadi}, \citenamefont {Golanbari}, \citenamefont {Enayati}, \citenamefont {Jalalzadeh}, \citenamefont {Nasri},\ and\ \citenamefont {Saaidi}}]{Mohammadi:2022fiv}%
  \BibitemOpen
  \bibfield  {author} {\bibinfo {author} {\bibfnamefont {A.}~\bibnamefont {Mohammadi}}, \bibinfo {author} {\bibfnamefont {T.}~\bibnamefont {Golanbari}}, \bibinfo {author} {\bibfnamefont {J.}~\bibnamefont {Enayati}}, \bibinfo {author} {\bibfnamefont {S.}~\bibnamefont {Jalalzadeh}}, \bibinfo {author} {\bibfnamefont {S.}~\bibnamefont {Nasri}},\ and\ \bibinfo {author} {\bibfnamefont {K.}~\bibnamefont {Saaidi}},\ }\bibfield  {title} {\bibinfo {title} {{Swampland criteria and reheating predictions in scalar\textendash{}tensor inflation}},\ }\href {https://doi.org/10.1142/S0218271822500791} {\bibfield  {journal} {\bibinfo  {journal} {Int. J. Mod. Phys. D}\ }\textbf {\bibinfo {volume} {31}},\ \bibinfo {pages} {2250079} (\bibinfo {year} {2022}{\natexlab{b}})}\BibitemShut {NoStop}%
\bibitem [{\citenamefont {Louis}\ \emph {et~al.}(2025)\citenamefont {Louis} \emph {et~al.}}]{ACT:2025fju}%
  \BibitemOpen
  \bibfield  {author} {\bibinfo {author} {\bibfnamefont {T.}~\bibnamefont {Louis}} \emph {et~al.} (\bibinfo {collaboration} {ACT}),\ }\bibfield  {title} {\bibinfo {title} {{The Atacama Cosmology Telescope: DR6 Power Spectra, Likelihoods and $\Lambda$CDM Parameters}},\ }\href@noop {} {\bibfield  {journal} {\bibinfo  {journal} {{arXiv preprint}}\ } (\bibinfo {year} {2025})},\ \Eprint {https://arxiv.org/abs/2503.14452} {arXiv:2503.14452 [astro-ph.CO]} \BibitemShut {NoStop}%
\bibitem [{\citenamefont {Calabrese}\ \emph {et~al.}(2025)\citenamefont {Calabrese} \emph {et~al.}}]{ACT:2025tim}%
  \BibitemOpen
  \bibfield  {author} {\bibinfo {author} {\bibfnamefont {E.}~\bibnamefont {Calabrese}} \emph {et~al.} (\bibinfo {collaboration} {ACT}),\ }\bibfield  {title} {\bibinfo {title} {{The Atacama Cosmology Telescope: DR6 Constraints on Extended Cosmological Models}},\ }\href@noop {} {\bibfield  {journal} {\bibinfo  {journal} {{arXiv preprint}}\ }\textbf {\bibinfo {volume} {{N/A}}},\ \bibinfo {pages} {{N/A}} (\bibinfo {year} {2025})},\ \Eprint {https://arxiv.org/abs/2503.14454} {arXiv:2503.14454 [astro-ph.CO]} \BibitemShut {NoStop}%
\bibitem [{\citenamefont {Kallosh}\ \emph {et~al.}(2025)\citenamefont {Kallosh}, \citenamefont {Linde},\ and\ \citenamefont {Roest}}]{Kallosh:2025rni}%
  \BibitemOpen
  \bibfield  {author} {\bibinfo {author} {\bibfnamefont {R.}~\bibnamefont {Kallosh}}, \bibinfo {author} {\bibfnamefont {A.}~\bibnamefont {Linde}},\ and\ \bibinfo {author} {\bibfnamefont {D.}~\bibnamefont {Roest}},\ }\bibfield  {title} {\bibinfo {title} {{A simple scenario for the last ACT}},\ }\href@noop {} {\  (\bibinfo {year} {2025})},\ \Eprint {https://arxiv.org/abs/2503.21030} {arXiv:2503.21030 [hep-th]} \BibitemShut {NoStop}%
\bibitem [{\citenamefont {Aoki}\ \emph {et~al.}(2025)\citenamefont {Aoki}, \citenamefont {Otsuka},\ and\ \citenamefont {Yanagita}}]{Aoki:2025wld}%
  \BibitemOpen
  \bibfield  {author} {\bibinfo {author} {\bibfnamefont {S.}~\bibnamefont {Aoki}}, \bibinfo {author} {\bibfnamefont {H.}~\bibnamefont {Otsuka}},\ and\ \bibinfo {author} {\bibfnamefont {R.}~\bibnamefont {Yanagita}},\ }\bibfield  {title} {\bibinfo {title} {{Higgs-Modular Inflation}},\ }\href@noop {} {\  (\bibinfo {year} {2025})},\ \Eprint {https://arxiv.org/abs/2504.01622} {arXiv:2504.01622 [hep-ph]} \BibitemShut {NoStop}%
\bibitem [{\citenamefont {Dioguardi}\ \emph {et~al.}(2025)\citenamefont {Dioguardi}, \citenamefont {Iovino},\ and\ \citenamefont {Racioppi}}]{Dioguardi:2025vci}%
  \BibitemOpen
  \bibfield  {author} {\bibinfo {author} {\bibfnamefont {C.}~\bibnamefont {Dioguardi}}, \bibinfo {author} {\bibfnamefont {A.~J.}\ \bibnamefont {Iovino}},\ and\ \bibinfo {author} {\bibfnamefont {A.}~\bibnamefont {Racioppi}},\ }\bibfield  {title} {\bibinfo {title} {{Fractional attractors in light of the latest ACT observations}},\ }\href@noop {} {\  (\bibinfo {year} {2025})},\ \Eprint {https://arxiv.org/abs/2504.02809} {arXiv:2504.02809 [gr-qc]} \BibitemShut {NoStop}%
\bibitem [{\citenamefont {Salvio}(2025)}]{Salvio:2025izr}%
  \BibitemOpen
  \bibfield  {author} {\bibinfo {author} {\bibfnamefont {A.}~\bibnamefont {Salvio}},\ }\bibfield  {title} {\bibinfo {title} {{Independent connection in ACTion during inflation}},\ }\href@noop {} {\  (\bibinfo {year} {2025})},\ \Eprint {https://arxiv.org/abs/2504.10488} {arXiv:2504.10488 [hep-ph]} \BibitemShut {NoStop}%
\bibitem [{\citenamefont {Brahma}\ and\ \citenamefont {Calder\'on-Figueroa}(2025)}]{Brahma:2025dio}%
  \BibitemOpen
  \bibfield  {author} {\bibinfo {author} {\bibfnamefont {S.}~\bibnamefont {Brahma}}\ and\ \bibinfo {author} {\bibfnamefont {J.}~\bibnamefont {Calder\'on-Figueroa}},\ }\bibfield  {title} {\bibinfo {title} {{Is the CMB revealing signs of pre-inflationary physics?}},\ }\href@noop {} {\  (\bibinfo {year} {2025})},\ \Eprint {https://arxiv.org/abs/2504.02746} {arXiv:2504.02746 [astro-ph.CO]} \BibitemShut {NoStop}%
\bibitem [{\citenamefont {Gao}\ \emph {et~al.}(2025)\citenamefont {Gao}, \citenamefont {Gong}, \citenamefont {Yi},\ and\ \citenamefont {Zhang}}]{Gao:2025onc}%
  \BibitemOpen
  \bibfield  {author} {\bibinfo {author} {\bibfnamefont {Q.}~\bibnamefont {Gao}}, \bibinfo {author} {\bibfnamefont {Y.}~\bibnamefont {Gong}}, \bibinfo {author} {\bibfnamefont {Z.}~\bibnamefont {Yi}},\ and\ \bibinfo {author} {\bibfnamefont {F.}~\bibnamefont {Zhang}},\ }\bibfield  {title} {\bibinfo {title} {{Non-minimal coupling in light of ACT}},\ }\href@noop {} {\  (\bibinfo {year} {2025})},\ \Eprint {https://arxiv.org/abs/2504.15218} {arXiv:2504.15218 [astro-ph.CO]} \BibitemShut {NoStop}%
\bibitem [{\citenamefont {Drees}\ and\ \citenamefont {Xu}(2025)}]{Drees:2025ngb}%
  \BibitemOpen
  \bibfield  {author} {\bibinfo {author} {\bibfnamefont {M.}~\bibnamefont {Drees}}\ and\ \bibinfo {author} {\bibfnamefont {Y.}~\bibnamefont {Xu}},\ }\bibfield  {title} {\bibinfo {title} {{Refined Predictions for Starobinsky Inflation and Post-inflationary Constraints in Light of ACT}},\ }\href@noop {} {\  (\bibinfo {year} {2025})},\ \Eprint {https://arxiv.org/abs/2504.20757} {arXiv:2504.20757 [astro-ph.CO]} \BibitemShut {NoStop}%
\bibitem [{\citenamefont {Zharov}\ \emph {et~al.}(2025)\citenamefont {Zharov}, \citenamefont {Sobol},\ and\ \citenamefont {Vilchinskii}}]{Zharov:2025evb}%
  \BibitemOpen
  \bibfield  {author} {\bibinfo {author} {\bibfnamefont {D.~S.}\ \bibnamefont {Zharov}}, \bibinfo {author} {\bibfnamefont {O.~O.}\ \bibnamefont {Sobol}},\ and\ \bibinfo {author} {\bibfnamefont {S.~I.}\ \bibnamefont {Vilchinskii}},\ }\bibfield  {title} {\bibinfo {title} {{Reheating ACTs on Starobinsky and Higgs inflation}},\ }\href@noop {} {\  (\bibinfo {year} {2025})},\ \Eprint {https://arxiv.org/abs/2505.01129} {arXiv:2505.01129 [astro-ph.CO]} \BibitemShut {NoStop}%
\bibitem [{\citenamefont {Yin}(2025)}]{Yin:2025rrs}%
  \BibitemOpen
  \bibfield  {author} {\bibinfo {author} {\bibfnamefont {W.}~\bibnamefont {Yin}},\ }\bibfield  {title} {\bibinfo {title} {{Higgs-like inflation under ACTivated mass}},\ }\href@noop {} {\  (\bibinfo {year} {2025})},\ \Eprint {https://arxiv.org/abs/2505.03004} {arXiv:2505.03004 [hep-ph]} \BibitemShut {NoStop}%
\bibitem [{\citenamefont {Liu}\ \emph {et~al.}(2025)\citenamefont {Liu}, \citenamefont {Yi},\ and\ \citenamefont {Gong}}]{Liu:2025qca}%
  \BibitemOpen
  \bibfield  {author} {\bibinfo {author} {\bibfnamefont {L.}~\bibnamefont {Liu}}, \bibinfo {author} {\bibfnamefont {Z.}~\bibnamefont {Yi}},\ and\ \bibinfo {author} {\bibfnamefont {Y.}~\bibnamefont {Gong}},\ }\bibfield  {title} {\bibinfo {title} {{Reconciling Higgs Inflation with ACT Observations through Reheating}},\ }\href@noop {} {\  (\bibinfo {year} {2025})},\ \Eprint {https://arxiv.org/abs/2505.02407} {arXiv:2505.02407 [astro-ph.CO]} \BibitemShut {NoStop}%
\bibitem [{\citenamefont {Gialamas}\ \emph {et~al.}(2025)\citenamefont {Gialamas}, \citenamefont {Katsoulas},\ and\ \citenamefont {Tamvakis}}]{Gialamas:2025ofz}%
  \BibitemOpen
  \bibfield  {author} {\bibinfo {author} {\bibfnamefont {I.~D.}\ \bibnamefont {Gialamas}}, \bibinfo {author} {\bibfnamefont {T.}~\bibnamefont {Katsoulas}},\ and\ \bibinfo {author} {\bibfnamefont {K.}~\bibnamefont {Tamvakis}},\ }\bibfield  {title} {\bibinfo {title} {{Keeping the relation between the Starobinsky model and no-scale supergravity ACTive}},\ }\href@noop {} {\  (\bibinfo {year} {2025})},\ \Eprint {https://arxiv.org/abs/2505.03608} {arXiv:2505.03608 [gr-qc]} \BibitemShut {NoStop}%
\bibitem [{\citenamefont {Kofman}\ \emph {et~al.}(1994)\citenamefont {Kofman}, \citenamefont {Linde},\ and\ \citenamefont {Starobinsky}}]{Kofman:1994rk}%
  \BibitemOpen
  \bibfield  {author} {\bibinfo {author} {\bibfnamefont {L.}~\bibnamefont {Kofman}}, \bibinfo {author} {\bibfnamefont {A.~D.}\ \bibnamefont {Linde}},\ and\ \bibinfo {author} {\bibfnamefont {A.~A.}\ \bibnamefont {Starobinsky}},\ }\bibfield  {title} {\bibinfo {title} {{Reheating after inflation}},\ }\href {https://doi.org/10.1103/PhysRevLett.73.3195} {\bibfield  {journal} {\bibinfo  {journal} {Phys. Rev. Lett.}\ }\textbf {\bibinfo {volume} {73}},\ \bibinfo {pages} {3195} (\bibinfo {year} {1994})},\ \Eprint {https://arxiv.org/abs/hep-th/9405187} {arXiv:hep-th/9405187} \BibitemShut {NoStop}%
\bibitem [{\citenamefont {Lozanov}(2019)}]{Lozanov:2019jxc}%
  \BibitemOpen
  \bibfield  {author} {\bibinfo {author} {\bibfnamefont {K.~D.}\ \bibnamefont {Lozanov}},\ }\bibfield  {title} {\bibinfo {title} {{Lectures on Reheating after Inflation}},\ }\href@noop {} {\  (\bibinfo {year} {2019})},\ \Eprint {https://arxiv.org/abs/1907.04402} {arXiv:1907.04402 [astro-ph.CO]} \BibitemShut {NoStop}%
\bibitem [{\citenamefont {Kofman}(1997)}]{Kofman:1997pt}%
  \BibitemOpen
  \bibfield  {author} {\bibinfo {author} {\bibfnamefont {L.}~\bibnamefont {Kofman}},\ }\bibfield  {title} {\bibinfo {title} {{Reheating and preheating after inflation}},\ }in\ \href@noop {} {\emph {\bibinfo {booktitle} {{3rd RESCEU International Symposium on Particle Cosmology}}}}\ (\bibinfo {year} {1997})\ pp.\ \bibinfo {pages} {1--8},\ \Eprint {https://arxiv.org/abs/hep-ph/9802285} {arXiv:hep-ph/9802285} \BibitemShut {NoStop}%
\bibitem [{\citenamefont {Bezrukov}\ and\ \citenamefont {Shaposhnikov}(2008)}]{Bezrukov:2007ep}%
  \BibitemOpen
  \bibfield  {author} {\bibinfo {author} {\bibfnamefont {F.~L.}\ \bibnamefont {Bezrukov}}\ and\ \bibinfo {author} {\bibfnamefont {M.}~\bibnamefont {Shaposhnikov}},\ }\bibfield  {title} {\bibinfo {title} {{The Standard Model Higgs boson as the inflaton}},\ }\href {https://doi.org/10.1016/j.physletb.2007.11.072} {\bibfield  {journal} {\bibinfo  {journal} {Phys. Lett. B}\ }\textbf {\bibinfo {volume} {659}},\ \bibinfo {pages} {703} (\bibinfo {year} {2008})},\ \Eprint {https://arxiv.org/abs/0710.3755} {arXiv:0710.3755 [hep-th]} \BibitemShut {NoStop}%
\bibitem [{\citenamefont {Bezrukov}\ \emph {et~al.}(2011)\citenamefont {Bezrukov}, \citenamefont {Magnin}, \citenamefont {Shaposhnikov},\ and\ \citenamefont {Sibiryakov}}]{Bezrukov:2010jz}%
  \BibitemOpen
  \bibfield  {author} {\bibinfo {author} {\bibfnamefont {F.}~\bibnamefont {Bezrukov}}, \bibinfo {author} {\bibfnamefont {A.}~\bibnamefont {Magnin}}, \bibinfo {author} {\bibfnamefont {M.}~\bibnamefont {Shaposhnikov}},\ and\ \bibinfo {author} {\bibfnamefont {S.}~\bibnamefont {Sibiryakov}},\ }\bibfield  {title} {\bibinfo {title} {{Higgs inflation: consistency and generalisations}},\ }\href {https://doi.org/10.1007/JHEP01(2011)016} {\bibfield  {journal} {\bibinfo  {journal} {JHEP}\ }\textbf {\bibinfo {volume} {01}},\ \bibinfo {pages} {016}},\ \Eprint {https://arxiv.org/abs/1008.5157} {arXiv:1008.5157 [hep-ph]} \BibitemShut {NoStop}%
\bibitem [{\citenamefont {Kallosh}\ and\ \citenamefont {Linde}(2013)}]{Kallosh:2013pby}%
  \BibitemOpen
  \bibfield  {author} {\bibinfo {author} {\bibfnamefont {R.}~\bibnamefont {Kallosh}}\ and\ \bibinfo {author} {\bibfnamefont {A.}~\bibnamefont {Linde}},\ }\bibfield  {title} {\bibinfo {title} {{Superconformal generalization of the chaotic inflation model $\frac{\lambda}{4} \phi^{4} - \frac{\xi}{2} \phi^{2}R$}},\ }\href {https://doi.org/10.1088/1475-7516/2013/06/027} {\bibfield  {journal} {\bibinfo  {journal} {JCAP}\ }\textbf {\bibinfo {volume} {06}},\ \bibinfo {pages} {027}},\ \Eprint {https://arxiv.org/abs/1306.3211} {arXiv:1306.3211 [hep-th]} \BibitemShut {NoStop}%
\bibitem [{\citenamefont {{Starobinsky}}(1982)}]{1982PhLB..117..175S}%
  \BibitemOpen
  \bibfield  {author} {\bibinfo {author} {\bibfnamefont {A.~A.}\ \bibnamefont {{Starobinsky}}},\ }\bibfield  {title} {\bibinfo {title} {{Dynamics of phase transition in the new inflationary universe scenario and generation of perturbations}},\ }\href {https://doi.org/10.1016/0370-2693(82)90541-X} {\bibfield  {journal} {\bibinfo  {journal} {Physics Letters B}\ }\textbf {\bibinfo {volume} {117}},\ \bibinfo {pages} {175} (\bibinfo {year} {1982})}\BibitemShut {NoStop}%
\bibitem [{\citenamefont {Starobinsky}(1983)}]{Starobinsky:1983zz}%
  \BibitemOpen
  \bibfield  {author} {\bibinfo {author} {\bibfnamefont {A.~A.}\ \bibnamefont {Starobinsky}},\ }\bibfield  {title} {\bibinfo {title} {{The Perturbation Spectrum Evolving from a Nonsingular Initially De-Sitter Cosmology and the Microwave Background Anisotropy}},\ }\href@noop {} {\bibfield  {journal} {\bibinfo  {journal} {Sov. Astron. Lett.}\ }\textbf {\bibinfo {volume} {9}},\ \bibinfo {pages} {302} (\bibinfo {year} {1983})}\BibitemShut {NoStop}%
\bibitem [{\citenamefont {Barvinsky}\ and\ \citenamefont {Kamenshchik}(1994)}]{Barvinsky:1994hx}%
  \BibitemOpen
  \bibfield  {author} {\bibinfo {author} {\bibfnamefont {A.~O.}\ \bibnamefont {Barvinsky}}\ and\ \bibinfo {author} {\bibfnamefont {A.~Y.}\ \bibnamefont {Kamenshchik}},\ }\bibfield  {title} {\bibinfo {title} {{Quantum scale of inflation and particle physics of the early universe}},\ }\href {https://doi.org/10.1016/0370-2693(94)91253-X} {\bibfield  {journal} {\bibinfo  {journal} {Phys. Lett. B}\ }\textbf {\bibinfo {volume} {332}},\ \bibinfo {pages} {270} (\bibinfo {year} {1994})},\ \Eprint {https://arxiv.org/abs/gr-qc/9404062} {arXiv:gr-qc/9404062} \BibitemShut {NoStop}%
\bibitem [{\citenamefont {Cervantes-Cota}\ and\ \citenamefont {Dehnen}(1995)}]{Cervantes-Cota:1995ehs}%
  \BibitemOpen
  \bibfield  {author} {\bibinfo {author} {\bibfnamefont {J.~L.}\ \bibnamefont {Cervantes-Cota}}\ and\ \bibinfo {author} {\bibfnamefont {H.}~\bibnamefont {Dehnen}},\ }\bibfield  {title} {\bibinfo {title} {{Induced gravity inflation in the standard model of particle physics}},\ }\href {https://doi.org/10.1016/0550-3213(95)00128-X} {\bibfield  {journal} {\bibinfo  {journal} {Nucl. Phys. B}\ }\textbf {\bibinfo {volume} {442}},\ \bibinfo {pages} {391} (\bibinfo {year} {1995})},\ \Eprint {https://arxiv.org/abs/astro-ph/9505069} {arXiv:astro-ph/9505069} \BibitemShut {NoStop}%
\bibitem [{\citenamefont {Barvinsky}\ \emph {et~al.}(2008)\citenamefont {Barvinsky}, \citenamefont {Kamenshchik},\ and\ \citenamefont {Starobinsky}}]{Barvinsky:2008ia}%
  \BibitemOpen
  \bibfield  {author} {\bibinfo {author} {\bibfnamefont {A.~O.}\ \bibnamefont {Barvinsky}}, \bibinfo {author} {\bibfnamefont {A.~Y.}\ \bibnamefont {Kamenshchik}},\ and\ \bibinfo {author} {\bibfnamefont {A.~A.}\ \bibnamefont {Starobinsky}},\ }\bibfield  {title} {\bibinfo {title} {{Inflation scenario via the Standard Model Higgs boson and LHC}},\ }\href {https://doi.org/10.1088/1475-7516/2008/11/021} {\bibfield  {journal} {\bibinfo  {journal} {JCAP}\ }\textbf {\bibinfo {volume} {11}},\ \bibinfo {pages} {021}},\ \Eprint {https://arxiv.org/abs/0809.2104} {arXiv:0809.2104 [hep-ph]} \BibitemShut {NoStop}%
\bibitem [{\citenamefont {De~Simone}\ \emph {et~al.}(2009)\citenamefont {De~Simone}, \citenamefont {Hertzberg},\ and\ \citenamefont {Wilczek}}]{DeSimone:2008ei}%
  \BibitemOpen
  \bibfield  {author} {\bibinfo {author} {\bibfnamefont {A.}~\bibnamefont {De~Simone}}, \bibinfo {author} {\bibfnamefont {M.~P.}\ \bibnamefont {Hertzberg}},\ and\ \bibinfo {author} {\bibfnamefont {F.}~\bibnamefont {Wilczek}},\ }\bibfield  {title} {\bibinfo {title} {{Running Inflation in the Standard Model}},\ }\href {https://doi.org/10.1016/j.physletb.2009.05.054} {\bibfield  {journal} {\bibinfo  {journal} {Phys. Lett. B}\ }\textbf {\bibinfo {volume} {678}},\ \bibinfo {pages} {1} (\bibinfo {year} {2009})},\ \Eprint {https://arxiv.org/abs/0812.4946} {arXiv:0812.4946 [hep-ph]} \BibitemShut {NoStop}%
\bibitem [{\citenamefont {Gialamas}\ \emph {et~al.}(2020{\natexlab{a}})\citenamefont {Gialamas}, \citenamefont {Karam}, \citenamefont {Lykkas},\ and\ \citenamefont {Pappas}}]{Gialamas:2020vto}%
  \BibitemOpen
  \bibfield  {author} {\bibinfo {author} {\bibfnamefont {I.~D.}\ \bibnamefont {Gialamas}}, \bibinfo {author} {\bibfnamefont {A.}~\bibnamefont {Karam}}, \bibinfo {author} {\bibfnamefont {A.}~\bibnamefont {Lykkas}},\ and\ \bibinfo {author} {\bibfnamefont {T.~D.}\ \bibnamefont {Pappas}},\ }\bibfield  {title} {\bibinfo {title} {{Palatini-Higgs inflation with nonminimal derivative coupling}},\ }\href {https://doi.org/10.1103/PhysRevD.102.063522} {\bibfield  {journal} {\bibinfo  {journal} {Phys. Rev. D}\ }\textbf {\bibinfo {volume} {102}},\ \bibinfo {pages} {063522} (\bibinfo {year} {2020}{\natexlab{a}})},\ \Eprint {https://arxiv.org/abs/2008.06371} {arXiv:2008.06371 [gr-qc]} \BibitemShut {NoStop}%
\bibitem [{\citenamefont {Bezrukov}\ \emph {et~al.}(2009)\citenamefont {Bezrukov}, \citenamefont {Magnin},\ and\ \citenamefont {Shaposhnikov}}]{Bezrukov:2008ej}%
  \BibitemOpen
  \bibfield  {author} {\bibinfo {author} {\bibfnamefont {F.~L.}\ \bibnamefont {Bezrukov}}, \bibinfo {author} {\bibfnamefont {A.}~\bibnamefont {Magnin}},\ and\ \bibinfo {author} {\bibfnamefont {M.}~\bibnamefont {Shaposhnikov}},\ }\bibfield  {title} {\bibinfo {title} {{Standard Model Higgs boson mass from inflation}},\ }\href {https://doi.org/10.1016/j.physletb.2009.03.035} {\bibfield  {journal} {\bibinfo  {journal} {Phys. Lett. B}\ }\textbf {\bibinfo {volume} {675}},\ \bibinfo {pages} {88} (\bibinfo {year} {2009})},\ \Eprint {https://arxiv.org/abs/0812.4950} {arXiv:0812.4950 [hep-ph]} \BibitemShut {NoStop}%
\bibitem [{\citenamefont {Barvinsky}\ \emph {et~al.}(2012)\citenamefont {Barvinsky}, \citenamefont {Kamenshchik}, \citenamefont {Kiefer}, \citenamefont {Starobinsky},\ and\ \citenamefont {Steinwachs}}]{Barvinsky:2009ii}%
  \BibitemOpen
  \bibfield  {author} {\bibinfo {author} {\bibfnamefont {A.~O.}\ \bibnamefont {Barvinsky}}, \bibinfo {author} {\bibfnamefont {A.~Y.}\ \bibnamefont {Kamenshchik}}, \bibinfo {author} {\bibfnamefont {C.}~\bibnamefont {Kiefer}}, \bibinfo {author} {\bibfnamefont {A.~A.}\ \bibnamefont {Starobinsky}},\ and\ \bibinfo {author} {\bibfnamefont {C.~F.}\ \bibnamefont {Steinwachs}},\ }\bibfield  {title} {\bibinfo {title} {{Higgs boson, renormalization group, and naturalness in cosmology}},\ }\href {https://doi.org/10.1140/epjc/s10052-012-2219-3} {\bibfield  {journal} {\bibinfo  {journal} {Eur. Phys. J. C}\ }\textbf {\bibinfo {volume} {72}},\ \bibinfo {pages} {2219} (\bibinfo {year} {2012})},\ \Eprint {https://arxiv.org/abs/0910.1041} {arXiv:0910.1041 [hep-ph]} \BibitemShut {NoStop}%
\bibitem [{\citenamefont {Rubio}(2019)}]{Rubio:2018ogq}%
  \BibitemOpen
  \bibfield  {author} {\bibinfo {author} {\bibfnamefont {J.}~\bibnamefont {Rubio}},\ }\bibfield  {title} {\bibinfo {title} {{Higgs inflation}},\ }\href {https://doi.org/10.3389/fspas.2018.00050} {\bibfield  {journal} {\bibinfo  {journal} {Front. Astron. Space Sci.}\ }\textbf {\bibinfo {volume} {5}},\ \bibinfo {pages} {50} (\bibinfo {year} {2019})},\ \Eprint {https://arxiv.org/abs/1807.02376} {arXiv:1807.02376 [hep-ph]} \BibitemShut {NoStop}%
\bibitem [{\citenamefont {Koshelev}\ \emph {et~al.}(2020)\citenamefont {Koshelev}, \citenamefont {Kumar},\ and\ \citenamefont {Starobinsky}}]{Koshelev:2020xby}%
  \BibitemOpen
  \bibfield  {author} {\bibinfo {author} {\bibfnamefont {A.~S.}\ \bibnamefont {Koshelev}}, \bibinfo {author} {\bibfnamefont {K.~S.}\ \bibnamefont {Kumar}},\ and\ \bibinfo {author} {\bibfnamefont {A.~A.}\ \bibnamefont {Starobinsky}},\ }\bibfield  {title} {\bibinfo {title} {{Analytic infinite derivative gravity, $R^2$-like inflation, quantum gravity and CMB}},\ }\href {https://doi.org/10.1142/S021827182043018X} {\bibfield  {journal} {\bibinfo  {journal} {Int. J. Mod. Phys. D}\ }\textbf {\bibinfo {volume} {29}},\ \bibinfo {pages} {2043018} (\bibinfo {year} {2020})},\ \Eprint {https://arxiv.org/abs/2005.09550} {arXiv:2005.09550 [hep-th]} \BibitemShut {NoStop}%
\bibitem [{\citenamefont {Gialamas}\ \emph {et~al.}(2023)\citenamefont {Gialamas}, \citenamefont {Karam}, \citenamefont {Pappas},\ and\ \citenamefont {Tomberg}}]{Gialamas:2023flv}%
  \BibitemOpen
  \bibfield  {author} {\bibinfo {author} {\bibfnamefont {I.~D.}\ \bibnamefont {Gialamas}}, \bibinfo {author} {\bibfnamefont {A.}~\bibnamefont {Karam}}, \bibinfo {author} {\bibfnamefont {T.~D.}\ \bibnamefont {Pappas}},\ and\ \bibinfo {author} {\bibfnamefont {E.}~\bibnamefont {Tomberg}},\ }\bibfield  {title} {\bibinfo {title} {{Implications of Palatini gravity for inflation and beyond}},\ }\href {https://doi.org/10.1142/S0219887823300076} {\bibfield  {journal} {\bibinfo  {journal} {Int. J. Geom. Meth. Mod. Phys.}\ }\textbf {\bibinfo {volume} {20}},\ \bibinfo {pages} {2330007} (\bibinfo {year} {2023})},\ \Eprint {https://arxiv.org/abs/2303.14148} {arXiv:2303.14148 [gr-qc]} \BibitemShut {NoStop}%
\bibitem [{\citenamefont {Gialamas}\ and\ \citenamefont {Tamvakis}(2023)}]{Gialamas:2022xtt}%
  \BibitemOpen
  \bibfield  {author} {\bibinfo {author} {\bibfnamefont {I.~D.}\ \bibnamefont {Gialamas}}\ and\ \bibinfo {author} {\bibfnamefont {K.}~\bibnamefont {Tamvakis}},\ }\bibfield  {title} {\bibinfo {title} {{Inflation in metric-affine quadratic gravity}},\ }\href {https://doi.org/10.1088/1475-7516/2023/03/042} {\bibfield  {journal} {\bibinfo  {journal} {JCAP}\ }\textbf {\bibinfo {volume} {03}},\ \bibinfo {pages} {042}},\ \Eprint {https://arxiv.org/abs/2212.09896} {arXiv:2212.09896 [gr-qc]} \BibitemShut {NoStop}%
\bibitem [{\citenamefont {Gialamas}\ \emph {et~al.}(2021)\citenamefont {Gialamas}, \citenamefont {Karam}, \citenamefont {Pappas},\ and\ \citenamefont {Spanos}}]{Gialamas:2021enw}%
  \BibitemOpen
  \bibfield  {author} {\bibinfo {author} {\bibfnamefont {I.~D.}\ \bibnamefont {Gialamas}}, \bibinfo {author} {\bibfnamefont {A.}~\bibnamefont {Karam}}, \bibinfo {author} {\bibfnamefont {T.~D.}\ \bibnamefont {Pappas}},\ and\ \bibinfo {author} {\bibfnamefont {V.~C.}\ \bibnamefont {Spanos}},\ }\bibfield  {title} {\bibinfo {title} {{Scale-invariant quadratic gravity and inflation in the Palatini formalism}},\ }\href {https://doi.org/10.1103/PhysRevD.104.023521} {\bibfield  {journal} {\bibinfo  {journal} {Phys. Rev. D}\ }\textbf {\bibinfo {volume} {104}},\ \bibinfo {pages} {023521} (\bibinfo {year} {2021})},\ \Eprint {https://arxiv.org/abs/2104.04550} {arXiv:2104.04550 [astro-ph.CO]} \BibitemShut {NoStop}%
\bibitem [{\citenamefont {Gialamas}\ \emph {et~al.}(2020{\natexlab{b}})\citenamefont {Gialamas}, \citenamefont {Karam},\ and\ \citenamefont {Racioppi}}]{Gialamas:2020snr}%
  \BibitemOpen
  \bibfield  {author} {\bibinfo {author} {\bibfnamefont {I.~D.}\ \bibnamefont {Gialamas}}, \bibinfo {author} {\bibfnamefont {A.}~\bibnamefont {Karam}},\ and\ \bibinfo {author} {\bibfnamefont {A.}~\bibnamefont {Racioppi}},\ }\bibfield  {title} {\bibinfo {title} {{Dynamically induced Planck scale and inflation in the Palatini formulation}},\ }\href {https://doi.org/10.1088/1475-7516/2020/11/014} {\bibfield  {journal} {\bibinfo  {journal} {JCAP}\ }\textbf {\bibinfo {volume} {11}},\ \bibinfo {pages} {014}},\ \Eprint {https://arxiv.org/abs/2006.09124} {arXiv:2006.09124 [gr-qc]} \BibitemShut {NoStop}%
\bibitem [{\citenamefont {Kim}\ \emph {et~al.}(2025{\natexlab{a}})\citenamefont {Kim}, \citenamefont {Yang},\ and\ \citenamefont {Zhang}}]{Kim:2025ikw}%
  \BibitemOpen
  \bibfield  {author} {\bibinfo {author} {\bibfnamefont {J.}~\bibnamefont {Kim}}, \bibinfo {author} {\bibfnamefont {Z.}~\bibnamefont {Yang}},\ and\ \bibinfo {author} {\bibfnamefont {Y.-l.}\ \bibnamefont {Zhang}},\ }\bibfield  {title} {\bibinfo {title} {{Gravitational Wave Signatures of Preheating in Higgs--$R^2$ Inflation}},\ }\href@noop {} {\  (\bibinfo {year} {2025}{\natexlab{a}})},\ \Eprint {https://arxiv.org/abs/2503.16907} {arXiv:2503.16907 [astro-ph.CO]} \BibitemShut {NoStop}%
\bibitem [{\citenamefont {Kim}\ \emph {et~al.}(2025{\natexlab{b}})\citenamefont {Kim}, \citenamefont {Wang}, \citenamefont {Zhang},\ and\ \citenamefont {Ren}}]{Kim:2025dyi}%
  \BibitemOpen
  \bibfield  {author} {\bibinfo {author} {\bibfnamefont {J.}~\bibnamefont {Kim}}, \bibinfo {author} {\bibfnamefont {X.}~\bibnamefont {Wang}}, \bibinfo {author} {\bibfnamefont {Y.-l.}\ \bibnamefont {Zhang}},\ and\ \bibinfo {author} {\bibfnamefont {Z.}~\bibnamefont {Ren}},\ }\bibfield  {title} {\bibinfo {title} {{Enhancement of primordial curvature perturbations in $R^3$-corrected Starobinsky-Higgs inflation}},\ }\href@noop {} {\  (\bibinfo {year} {2025}{\natexlab{b}})},\ \Eprint {https://arxiv.org/abs/2504.12035} {arXiv:2504.12035 [astro-ph.CO]} \BibitemShut {NoStop}%
\bibitem [{\citenamefont {van~de Bruck}\ and\ \citenamefont {Longden}(2016)}]{vandeBruck:2015gjd}%
  \BibitemOpen
  \bibfield  {author} {\bibinfo {author} {\bibfnamefont {C.}~\bibnamefont {van~de Bruck}}\ and\ \bibinfo {author} {\bibfnamefont {C.}~\bibnamefont {Longden}},\ }\bibfield  {title} {\bibinfo {title} {{Higgs Inflation with a Gauss-Bonnet term in the Jordan Frame}},\ }\href {https://doi.org/10.1103/PhysRevD.93.063519} {\bibfield  {journal} {\bibinfo  {journal} {Phys. Rev. D}\ }\textbf {\bibinfo {volume} {93}},\ \bibinfo {pages} {063519} (\bibinfo {year} {2016})},\ \Eprint {https://arxiv.org/abs/1512.04768} {arXiv:1512.04768 [hep-ph]} \BibitemShut {NoStop}%
\bibitem [{\citenamefont {Guo}\ and\ \citenamefont {Schwarz}(2009)}]{Guo:2009uk}%
  \BibitemOpen
  \bibfield  {author} {\bibinfo {author} {\bibfnamefont {Z.-K.}\ \bibnamefont {Guo}}\ and\ \bibinfo {author} {\bibfnamefont {D.~J.}\ \bibnamefont {Schwarz}},\ }\bibfield  {title} {\bibinfo {title} {{Power spectra from an inflaton coupled to the Gauss-Bonnet term}},\ }\href {https://doi.org/10.1103/PhysRevD.80.063523} {\bibfield  {journal} {\bibinfo  {journal} {Phys. Rev. D}\ }\textbf {\bibinfo {volume} {80}},\ \bibinfo {pages} {063523} (\bibinfo {year} {2009})},\ \Eprint {https://arxiv.org/abs/0907.0427} {arXiv:0907.0427 [hep-th]} \BibitemShut {NoStop}%
\bibitem [{\citenamefont {Guo}\ and\ \citenamefont {Schwarz}(2010)}]{Guo:2010jr}%
  \BibitemOpen
  \bibfield  {author} {\bibinfo {author} {\bibfnamefont {Z.-K.}\ \bibnamefont {Guo}}\ and\ \bibinfo {author} {\bibfnamefont {D.~J.}\ \bibnamefont {Schwarz}},\ }\bibfield  {title} {\bibinfo {title} {{Slow-roll inflation with a Gauss-Bonnet correction}},\ }\href {https://doi.org/10.1103/PhysRevD.81.123520} {\bibfield  {journal} {\bibinfo  {journal} {Phys. Rev. D}\ }\textbf {\bibinfo {volume} {81}},\ \bibinfo {pages} {123520} (\bibinfo {year} {2010})},\ \Eprint {https://arxiv.org/abs/1001.1897} {arXiv:1001.1897 [hep-th]} \BibitemShut {NoStop}%
\bibitem [{\citenamefont {Koh}\ \emph {et~al.}(2017)\citenamefont {Koh}, \citenamefont {Lee},\ and\ \citenamefont {Tumurtushaa}}]{Koh:2016abf}%
  \BibitemOpen
  \bibfield  {author} {\bibinfo {author} {\bibfnamefont {S.}~\bibnamefont {Koh}}, \bibinfo {author} {\bibfnamefont {B.-H.}\ \bibnamefont {Lee}},\ and\ \bibinfo {author} {\bibfnamefont {G.}~\bibnamefont {Tumurtushaa}},\ }\bibfield  {title} {\bibinfo {title} {{Reconstruction of the Scalar Field Potential in Inflationary Models with a Gauss-Bonnet term}},\ }\href {https://doi.org/10.1103/PhysRevD.95.123509} {\bibfield  {journal} {\bibinfo  {journal} {Phys. Rev. D}\ }\textbf {\bibinfo {volume} {95}},\ \bibinfo {pages} {123509} (\bibinfo {year} {2017})},\ \Eprint {https://arxiv.org/abs/1610.04360} {arXiv:1610.04360 [gr-qc]} \BibitemShut {NoStop}%
\bibitem [{\citenamefont {Pozdeeva}(2020)}]{Pozdeeva:2020shl}%
  \BibitemOpen
  \bibfield  {author} {\bibinfo {author} {\bibfnamefont {E.~O.}\ \bibnamefont {Pozdeeva}},\ }\bibfield  {title} {\bibinfo {title} {{Generalization of cosmological attractor approach to Einstein\textendash{}Gauss\textendash{}Bonnet gravity}},\ }\href {https://doi.org/10.1140/epjc/s10052-020-8176-3} {\bibfield  {journal} {\bibinfo  {journal} {Eur. Phys. J. C}\ }\textbf {\bibinfo {volume} {80}},\ \bibinfo {pages} {612} (\bibinfo {year} {2020})},\ \Eprint {https://arxiv.org/abs/2005.10133} {arXiv:2005.10133 [gr-qc]} \BibitemShut {NoStop}%
\bibitem [{\citenamefont {Satoh}\ and\ \citenamefont {Soda}(2008)}]{Satoh:2008ck}%
  \BibitemOpen
  \bibfield  {author} {\bibinfo {author} {\bibfnamefont {M.}~\bibnamefont {Satoh}}\ and\ \bibinfo {author} {\bibfnamefont {J.}~\bibnamefont {Soda}},\ }\bibfield  {title} {\bibinfo {title} {{Higher Curvature Corrections to Primordial Fluctuations in Slow-roll Inflation}},\ }\href {https://doi.org/10.1088/1475-7516/2008/09/019} {\bibfield  {journal} {\bibinfo  {journal} {JCAP}\ }\textbf {\bibinfo {volume} {09}},\ \bibinfo {pages} {019}},\ \Eprint {https://arxiv.org/abs/0806.4594} {arXiv:0806.4594 [astro-ph]} \BibitemShut {NoStop}%
\bibitem [{\citenamefont {Jiang}\ \emph {et~al.}(2013)\citenamefont {Jiang}, \citenamefont {Hu},\ and\ \citenamefont {Guo}}]{Jiang:2013gza}%
  \BibitemOpen
  \bibfield  {author} {\bibinfo {author} {\bibfnamefont {P.-X.}\ \bibnamefont {Jiang}}, \bibinfo {author} {\bibfnamefont {J.-W.}\ \bibnamefont {Hu}},\ and\ \bibinfo {author} {\bibfnamefont {Z.-K.}\ \bibnamefont {Guo}},\ }\bibfield  {title} {\bibinfo {title} {{Inflation coupled to a Gauss-Bonnet term}},\ }\href {https://doi.org/10.1103/PhysRevD.88.123508} {\bibfield  {journal} {\bibinfo  {journal} {Phys. Rev. D}\ }\textbf {\bibinfo {volume} {88}},\ \bibinfo {pages} {123508} (\bibinfo {year} {2013})},\ \Eprint {https://arxiv.org/abs/1310.5579} {arXiv:1310.5579 [hep-th]} \BibitemShut {NoStop}%
\bibitem [{\citenamefont {Koh}\ \emph {et~al.}(2014)\citenamefont {Koh}, \citenamefont {Lee}, \citenamefont {Lee},\ and\ \citenamefont {Tumurtushaa}}]{Koh:2014bka}%
  \BibitemOpen
  \bibfield  {author} {\bibinfo {author} {\bibfnamefont {S.}~\bibnamefont {Koh}}, \bibinfo {author} {\bibfnamefont {B.-H.}\ \bibnamefont {Lee}}, \bibinfo {author} {\bibfnamefont {W.}~\bibnamefont {Lee}},\ and\ \bibinfo {author} {\bibfnamefont {G.}~\bibnamefont {Tumurtushaa}},\ }\bibfield  {title} {\bibinfo {title} {{Observational constraints on slow-roll inflation coupled to a Gauss-Bonnet term}},\ }\href {https://doi.org/10.1103/PhysRevD.90.063527} {\bibfield  {journal} {\bibinfo  {journal} {Phys. Rev. D}\ }\textbf {\bibinfo {volume} {90}},\ \bibinfo {pages} {063527} (\bibinfo {year} {2014})},\ \Eprint {https://arxiv.org/abs/1404.6096} {arXiv:1404.6096 [gr-qc]} \BibitemShut {NoStop}%
\bibitem [{\citenamefont {Koh}\ \emph {et~al.}(2018)\citenamefont {Koh}, \citenamefont {Lee},\ and\ \citenamefont {Tumurtushaa}}]{Koh:2018qcy}%
  \BibitemOpen
  \bibfield  {author} {\bibinfo {author} {\bibfnamefont {S.}~\bibnamefont {Koh}}, \bibinfo {author} {\bibfnamefont {B.-H.}\ \bibnamefont {Lee}},\ and\ \bibinfo {author} {\bibfnamefont {G.}~\bibnamefont {Tumurtushaa}},\ }\bibfield  {title} {\bibinfo {title} {{Constraints on the reheating parameters after Gauss-Bonnet inflation from primordial gravitational waves}},\ }\href {https://doi.org/10.1103/PhysRevD.98.103511} {\bibfield  {journal} {\bibinfo  {journal} {Phys. Rev. D}\ }\textbf {\bibinfo {volume} {98}},\ \bibinfo {pages} {103511} (\bibinfo {year} {2018})},\ \Eprint {https://arxiv.org/abs/1807.04424} {arXiv:1807.04424 [astro-ph.CO]} \BibitemShut {NoStop}%
\bibitem [{\citenamefont {Mathew}\ and\ \citenamefont {Shankaranarayanan}(2016)}]{Mathew:2016anx}%
  \BibitemOpen
  \bibfield  {author} {\bibinfo {author} {\bibfnamefont {J.}~\bibnamefont {Mathew}}\ and\ \bibinfo {author} {\bibfnamefont {S.}~\bibnamefont {Shankaranarayanan}},\ }\bibfield  {title} {\bibinfo {title} {{Low scale Higgs inflation with Gauss\textendash{}Bonnet coupling}},\ }\href {https://doi.org/10.1016/j.astropartphys.2016.07.004} {\bibfield  {journal} {\bibinfo  {journal} {Astropart. Phys.}\ }\textbf {\bibinfo {volume} {84}},\ \bibinfo {pages} {1} (\bibinfo {year} {2016})},\ \Eprint {https://arxiv.org/abs/1602.00411} {arXiv:1602.00411 [astro-ph.CO]} \BibitemShut {NoStop}%
\bibitem [{\citenamefont {Pozdeeva}\ \emph {et~al.}(2020)\citenamefont {Pozdeeva}, \citenamefont {Gangopadhyay}, \citenamefont {Sami}, \citenamefont {Toporensky},\ and\ \citenamefont {Vernov}}]{Pozdeeva:2020apf}%
  \BibitemOpen
  \bibfield  {author} {\bibinfo {author} {\bibfnamefont {E.~O.}\ \bibnamefont {Pozdeeva}}, \bibinfo {author} {\bibfnamefont {M.~R.}\ \bibnamefont {Gangopadhyay}}, \bibinfo {author} {\bibfnamefont {M.}~\bibnamefont {Sami}}, \bibinfo {author} {\bibfnamefont {A.~V.}\ \bibnamefont {Toporensky}},\ and\ \bibinfo {author} {\bibfnamefont {S.~Y.}\ \bibnamefont {Vernov}},\ }\bibfield  {title} {\bibinfo {title} {{Inflation with a quartic potential in the framework of Einstein-Gauss-Bonnet gravity}},\ }\href {https://doi.org/10.1103/PhysRevD.102.043525} {\bibfield  {journal} {\bibinfo  {journal} {Phys. Rev. D}\ }\textbf {\bibinfo {volume} {102}},\ \bibinfo {pages} {043525} (\bibinfo {year} {2020})},\ \Eprint {https://arxiv.org/abs/2006.08027} {arXiv:2006.08027 [gr-qc]} \BibitemShut {NoStop}%
\bibitem [{\citenamefont {Pozdeeva}\ \emph {et~al.}(2016)\citenamefont {Pozdeeva}, \citenamefont {Skugoreva}, \citenamefont {Toporensky},\ and\ \citenamefont {Vernov}}]{Pozdeeva:2016cja}%
  \BibitemOpen
  \bibfield  {author} {\bibinfo {author} {\bibfnamefont {E.~O.}\ \bibnamefont {Pozdeeva}}, \bibinfo {author} {\bibfnamefont {M.~A.}\ \bibnamefont {Skugoreva}}, \bibinfo {author} {\bibfnamefont {A.~V.}\ \bibnamefont {Toporensky}},\ and\ \bibinfo {author} {\bibfnamefont {S.~Y.}\ \bibnamefont {Vernov}},\ }\bibfield  {title} {\bibinfo {title} {{Possible evolution of a bouncing universe in cosmological models with non-minimally coupled scalar fields}},\ }\href {https://doi.org/10.1088/1475-7516/2016/12/006} {\bibfield  {journal} {\bibinfo  {journal} {JCAP}\ }\textbf {\bibinfo {volume} {12}},\ \bibinfo {pages} {006}},\ \Eprint {https://arxiv.org/abs/1608.08214} {arXiv:1608.08214 [gr-qc]} \BibitemShut {NoStop}%
\bibitem [{\citenamefont {Nozari}\ and\ \citenamefont {Rashidi}(2017)}]{Nozari:2017rta}%
  \BibitemOpen
  \bibfield  {author} {\bibinfo {author} {\bibfnamefont {K.}~\bibnamefont {Nozari}}\ and\ \bibinfo {author} {\bibfnamefont {N.}~\bibnamefont {Rashidi}},\ }\bibfield  {title} {\bibinfo {title} {{Perturbation, non-Gaussianity, and reheating in a Gauss-Bonnet $\alpha$-attractor model}},\ }\href {https://doi.org/10.1103/PhysRevD.95.123518} {\bibfield  {journal} {\bibinfo  {journal} {Phys. Rev. D}\ }\textbf {\bibinfo {volume} {95}},\ \bibinfo {pages} {123518} (\bibinfo {year} {2017})},\ \Eprint {https://arxiv.org/abs/1705.02617} {arXiv:1705.02617 [astro-ph.CO]} \BibitemShut {NoStop}%
\bibitem [{\citenamefont {Yi}\ and\ \citenamefont {Gong}(2019)}]{Yi:2018dhl}%
  \BibitemOpen
  \bibfield  {author} {\bibinfo {author} {\bibfnamefont {Z.}~\bibnamefont {Yi}}\ and\ \bibinfo {author} {\bibfnamefont {Y.}~\bibnamefont {Gong}},\ }\bibfield  {title} {\bibinfo {title} {{Gauss\textendash{}Bonnet Inflation and the String Swampland}},\ }\href {https://doi.org/10.3390/universe5090200} {\bibfield  {journal} {\bibinfo  {journal} {Universe}\ }\textbf {\bibinfo {volume} {5}},\ \bibinfo {pages} {200} (\bibinfo {year} {2019})},\ \Eprint {https://arxiv.org/abs/1811.01625} {arXiv:1811.01625 [gr-qc]} \BibitemShut {NoStop}%
\bibitem [{\citenamefont {Odintsov}\ and\ \citenamefont {Oikonomou}(2018)}]{Odintsov:2018zhw}%
  \BibitemOpen
  \bibfield  {author} {\bibinfo {author} {\bibfnamefont {S.~D.}\ \bibnamefont {Odintsov}}\ and\ \bibinfo {author} {\bibfnamefont {V.~K.}\ \bibnamefont {Oikonomou}},\ }\bibfield  {title} {\bibinfo {title} {{Viable Inflation in Scalar-Gauss-Bonnet Gravity and Reconstruction from Observational Indices}},\ }\href {https://doi.org/10.1103/PhysRevD.98.044039} {\bibfield  {journal} {\bibinfo  {journal} {Phys. Rev. D}\ }\textbf {\bibinfo {volume} {98}},\ \bibinfo {pages} {044039} (\bibinfo {year} {2018})},\ \Eprint {https://arxiv.org/abs/1808.05045} {arXiv:1808.05045 [gr-qc]} \BibitemShut {NoStop}%
\bibitem [{\citenamefont {Fomin}\ and\ \citenamefont {Chervon}(2019)}]{Fomin:2019yls}%
  \BibitemOpen
  \bibfield  {author} {\bibinfo {author} {\bibfnamefont {I.~V.}\ \bibnamefont {Fomin}}\ and\ \bibinfo {author} {\bibfnamefont {S.~V.}\ \bibnamefont {Chervon}},\ }\bibfield  {title} {\bibinfo {title} {{Reconstruction of general relativistic cosmological solutions in modified gravity theories}},\ }\href {https://doi.org/10.1103/PhysRevD.100.023511} {\bibfield  {journal} {\bibinfo  {journal} {Phys. Rev. D}\ }\textbf {\bibinfo {volume} {100}},\ \bibinfo {pages} {023511} (\bibinfo {year} {2019})},\ \Eprint {https://arxiv.org/abs/1903.03974} {arXiv:1903.03974 [gr-qc]} \BibitemShut {NoStop}%
\bibitem [{\citenamefont {Fomin}(2020)}]{Fomin:2020hfh}%
  \BibitemOpen
  \bibfield  {author} {\bibinfo {author} {\bibfnamefont {I.}~\bibnamefont {Fomin}},\ }\bibfield  {title} {\bibinfo {title} {{Gauss\textendash{}Bonnet term corrections in scalar field cosmology}},\ }\href {https://doi.org/10.1140/epjc/s10052-020-08718-w} {\bibfield  {journal} {\bibinfo  {journal} {Eur. Phys. J. C}\ }\textbf {\bibinfo {volume} {80}},\ \bibinfo {pages} {1145} (\bibinfo {year} {2020})},\ \Eprint {https://arxiv.org/abs/2004.08065} {arXiv:2004.08065 [gr-qc]} \BibitemShut {NoStop}%
\bibitem [{\citenamefont {Kleidis}\ and\ \citenamefont {Oikonomou}(2019)}]{Kleidis:2019ywv}%
  \BibitemOpen
  \bibfield  {author} {\bibinfo {author} {\bibfnamefont {K.}~\bibnamefont {Kleidis}}\ and\ \bibinfo {author} {\bibfnamefont {V.~K.}\ \bibnamefont {Oikonomou}},\ }\bibfield  {title} {\bibinfo {title} {{A Study of an Einstein Gauss-Bonnet Quintessential Inflationary Model}},\ }\href {https://doi.org/10.1016/j.nuclphysb.2019.114765} {\bibfield  {journal} {\bibinfo  {journal} {Nucl. Phys. B}\ }\textbf {\bibinfo {volume} {948}},\ \bibinfo {pages} {114765} (\bibinfo {year} {2019})},\ \Eprint {https://arxiv.org/abs/1909.05318} {arXiv:1909.05318 [gr-qc]} \BibitemShut {NoStop}%
\bibitem [{\citenamefont {Odintsov}\ \emph {et~al.}(2020{\natexlab{a}})\citenamefont {Odintsov}, \citenamefont {Oikonomou},\ and\ \citenamefont {Fronimos}}]{Odintsov:2020sqy}%
  \BibitemOpen
  \bibfield  {author} {\bibinfo {author} {\bibfnamefont {S.~D.}\ \bibnamefont {Odintsov}}, \bibinfo {author} {\bibfnamefont {V.~K.}\ \bibnamefont {Oikonomou}},\ and\ \bibinfo {author} {\bibfnamefont {F.~P.}\ \bibnamefont {Fronimos}},\ }\bibfield  {title} {\bibinfo {title} {{Rectifying Einstein-Gauss-Bonnet Inflation in View of GW170817}},\ }\href {https://doi.org/10.1016/j.nuclphysb.2020.115135} {\bibfield  {journal} {\bibinfo  {journal} {Nucl. Phys. B}\ }\textbf {\bibinfo {volume} {958}},\ \bibinfo {pages} {115135} (\bibinfo {year} {2020}{\natexlab{a}})},\ \Eprint {https://arxiv.org/abs/2003.13724} {arXiv:2003.13724 [gr-qc]} \BibitemShut {NoStop}%
\bibitem [{\citenamefont {Odintsov}\ and\ \citenamefont {Oikonomou}(2020)}]{Odintsov:2020zkl}%
  \BibitemOpen
  \bibfield  {author} {\bibinfo {author} {\bibfnamefont {S.~D.}\ \bibnamefont {Odintsov}}\ and\ \bibinfo {author} {\bibfnamefont {V.~K.}\ \bibnamefont {Oikonomou}},\ }\bibfield  {title} {\bibinfo {title} {{Swampland implications of GW170817-compatible Einstein-Gauss-Bonnet gravity}},\ }\href {https://doi.org/10.1016/j.physletb.2020.135437} {\bibfield  {journal} {\bibinfo  {journal} {Phys. Lett. B}\ }\textbf {\bibinfo {volume} {805}},\ \bibinfo {pages} {135437} (\bibinfo {year} {2020})},\ \Eprint {https://arxiv.org/abs/2004.00479} {arXiv:2004.00479 [gr-qc]} \BibitemShut {NoStop}%
\bibitem [{\citenamefont {Kawai}\ and\ \citenamefont {Kim}(2021{\natexlab{a}})}]{Kawai:2021bye}%
  \BibitemOpen
  \bibfield  {author} {\bibinfo {author} {\bibfnamefont {S.}~\bibnamefont {Kawai}}\ and\ \bibinfo {author} {\bibfnamefont {J.}~\bibnamefont {Kim}},\ }\bibfield  {title} {\bibinfo {title} {{CMB from a Gauss-Bonnet-induced de Sitter fixed point}},\ }\href {https://doi.org/10.1103/PhysRevD.104.043525} {\bibfield  {journal} {\bibinfo  {journal} {Phys. Rev. D}\ }\textbf {\bibinfo {volume} {104}},\ \bibinfo {pages} {043525} (\bibinfo {year} {2021}{\natexlab{a}})},\ \Eprint {https://arxiv.org/abs/2105.04386} {arXiv:2105.04386 [hep-ph]} \BibitemShut {NoStop}%
\bibitem [{\citenamefont {Kawai}\ and\ \citenamefont {Kim}(2019)}]{Kawai:2017kqt}%
  \BibitemOpen
  \bibfield  {author} {\bibinfo {author} {\bibfnamefont {S.}~\bibnamefont {Kawai}}\ and\ \bibinfo {author} {\bibfnamefont {J.}~\bibnamefont {Kim}},\ }\bibfield  {title} {\bibinfo {title} {{Gauss\textendash{}Bonnet Chern\textendash{}Simons gravitational wave leptogenesis}},\ }\href {https://doi.org/10.1016/j.physletb.2018.12.019} {\bibfield  {journal} {\bibinfo  {journal} {Phys. Lett. B}\ }\textbf {\bibinfo {volume} {789}},\ \bibinfo {pages} {145} (\bibinfo {year} {2019})},\ \Eprint {https://arxiv.org/abs/1702.07689} {arXiv:1702.07689 [hep-th]} \BibitemShut {NoStop}%
\bibitem [{\citenamefont {Oikonomou}(2022)}]{Oikonomou:2022xoq}%
  \BibitemOpen
  \bibfield  {author} {\bibinfo {author} {\bibfnamefont {V.~K.}\ \bibnamefont {Oikonomou}},\ }\bibfield  {title} {\bibinfo {title} {{Primordial gravitational waves predictions for GW170817-compatible Einstein\textendash{}Gauss\textendash{}Bonnet theory}},\ }\href {https://doi.org/10.1016/j.astropartphys.2022.102718} {\bibfield  {journal} {\bibinfo  {journal} {Astropart. Phys.}\ }\textbf {\bibinfo {volume} {141}},\ \bibinfo {pages} {102718} (\bibinfo {year} {2022})},\ \Eprint {https://arxiv.org/abs/2204.06304} {arXiv:2204.06304 [gr-qc]} \BibitemShut {NoStop}%
\bibitem [{\citenamefont {Oikonomou}\ \emph {et~al.}(2022)\citenamefont {Oikonomou}, \citenamefont {Katzanis},\ and\ \citenamefont {Papadimitriou}}]{Oikonomou:2022ksx}%
  \BibitemOpen
  \bibfield  {author} {\bibinfo {author} {\bibfnamefont {V.~K.}\ \bibnamefont {Oikonomou}}, \bibinfo {author} {\bibfnamefont {P.~D.}\ \bibnamefont {Katzanis}},\ and\ \bibinfo {author} {\bibfnamefont {I.~C.}\ \bibnamefont {Papadimitriou}},\ }\bibfield  {title} {\bibinfo {title} {{Bottom-up reconstruction of viable GW170817 compatible Einstein\textendash{}Gauss\textendash{}Bonnet theories}},\ }\href {https://doi.org/10.1088/1361-6382/ac5eba} {\bibfield  {journal} {\bibinfo  {journal} {Class. Quant. Grav.}\ }\textbf {\bibinfo {volume} {39}},\ \bibinfo {pages} {095008} (\bibinfo {year} {2022})},\ \Eprint {https://arxiv.org/abs/2203.09867} {arXiv:2203.09867 [gr-qc]} \BibitemShut {NoStop}%
\bibitem [{\citenamefont {Cognola}\ \emph {et~al.}(2007)\citenamefont {Cognola}, \citenamefont {Elizalde}, \citenamefont {Nojiri}, \citenamefont {Odintsov},\ and\ \citenamefont {Zerbini}}]{Cognola:2006sp}%
  \BibitemOpen
  \bibfield  {author} {\bibinfo {author} {\bibfnamefont {G.}~\bibnamefont {Cognola}}, \bibinfo {author} {\bibfnamefont {E.}~\bibnamefont {Elizalde}}, \bibinfo {author} {\bibfnamefont {S.}~\bibnamefont {Nojiri}}, \bibinfo {author} {\bibfnamefont {S.}~\bibnamefont {Odintsov}},\ and\ \bibinfo {author} {\bibfnamefont {S.}~\bibnamefont {Zerbini}},\ }\bibfield  {title} {\bibinfo {title} {{String-inspired Gauss-Bonnet gravity reconstructed from the universe expansion history and yielding the transition from matter dominance to dark energy}},\ }\href {https://doi.org/10.1103/PhysRevD.75.086002} {\bibfield  {journal} {\bibinfo  {journal} {Phys. Rev. D}\ }\textbf {\bibinfo {volume} {75}},\ \bibinfo {pages} {086002} (\bibinfo {year} {2007})},\ \Eprint {https://arxiv.org/abs/hep-th/0611198} {arXiv:hep-th/0611198} \BibitemShut {NoStop}%
\bibitem [{\citenamefont {Odintsov}\ \emph {et~al.}(2020{\natexlab{b}})\citenamefont {Odintsov}, \citenamefont {Oikonomou},\ and\ \citenamefont {Fronimos}}]{Odintsov:2020xji}%
  \BibitemOpen
  \bibfield  {author} {\bibinfo {author} {\bibfnamefont {S.~D.}\ \bibnamefont {Odintsov}}, \bibinfo {author} {\bibfnamefont {V.~K.}\ \bibnamefont {Oikonomou}},\ and\ \bibinfo {author} {\bibfnamefont {F.~P.}\ \bibnamefont {Fronimos}},\ }\bibfield  {title} {\bibinfo {title} {{Non-minimally coupled Einstein-Gauss-Bonnet inflation phenomenology in view of GW170817}},\ }\href {https://doi.org/10.1016/j.aop.2020.168250} {\bibfield  {journal} {\bibinfo  {journal} {Annals Phys.}\ }\textbf {\bibinfo {volume} {420}},\ \bibinfo {pages} {168250} (\bibinfo {year} {2020}{\natexlab{b}})},\ \Eprint {https://arxiv.org/abs/2007.02309} {arXiv:2007.02309 [gr-qc]} \BibitemShut {NoStop}%
\bibitem [{\citenamefont {Odintsov}\ \emph {et~al.}(2020{\natexlab{c}})\citenamefont {Odintsov}, \citenamefont {Oikonomou}, \citenamefont {Fronimos},\ and\ \citenamefont {Venikoudis}}]{Odintsov:2020mkz}%
  \BibitemOpen
  \bibfield  {author} {\bibinfo {author} {\bibfnamefont {S.~D.}\ \bibnamefont {Odintsov}}, \bibinfo {author} {\bibfnamefont {V.~K.}\ \bibnamefont {Oikonomou}}, \bibinfo {author} {\bibfnamefont {F.~P.}\ \bibnamefont {Fronimos}},\ and\ \bibinfo {author} {\bibfnamefont {S.~A.}\ \bibnamefont {Venikoudis}},\ }\bibfield  {title} {\bibinfo {title} {{GW170817-compatible constant-roll Einstein\textendash{}Gauss\textendash{}Bonnet inflation and non-Gaussianities}},\ }\href {https://doi.org/10.1016/j.dark.2020.100718} {\bibfield  {journal} {\bibinfo  {journal} {Phys. Dark Univ.}\ }\textbf {\bibinfo {volume} {30}},\ \bibinfo {pages} {100718} (\bibinfo {year} {2020}{\natexlab{c}})},\ \Eprint {https://arxiv.org/abs/2009.06113} {arXiv:2009.06113 [gr-qc]} \BibitemShut {NoStop}%
\bibitem [{\citenamefont {Oikonomou}\ and\ \citenamefont {Fronimos}(2021)}]{Oikonomou:2020sij}%
  \BibitemOpen
  \bibfield  {author} {\bibinfo {author} {\bibfnamefont {V.~K.}\ \bibnamefont {Oikonomou}}\ and\ \bibinfo {author} {\bibfnamefont {F.~P.}\ \bibnamefont {Fronimos}},\ }\bibfield  {title} {\bibinfo {title} {{Reviving non-minimal Horndeski-like theories after GW170817: kinetic coupling corrected Einstein\textendash{}Gauss\textendash{}Bonnet inflation}},\ }\href {https://doi.org/10.1088/1361-6382/abce47} {\bibfield  {journal} {\bibinfo  {journal} {Class. Quant. Grav.}\ }\textbf {\bibinfo {volume} {38}},\ \bibinfo {pages} {035013} (\bibinfo {year} {2021})},\ \Eprint {https://arxiv.org/abs/2006.05512} {arXiv:2006.05512 [gr-qc]} \BibitemShut {NoStop}%
\bibitem [{\citenamefont {Nojiri}\ \emph {et~al.}(2019)\citenamefont {Nojiri}, \citenamefont {Odintsov}, \citenamefont {Oikonomou}, \citenamefont {Chatzarakis},\ and\ \citenamefont {Paul}}]{Nojiri:2019dwl}%
  \BibitemOpen
  \bibfield  {author} {\bibinfo {author} {\bibfnamefont {S.}~\bibnamefont {Nojiri}}, \bibinfo {author} {\bibfnamefont {S.~D.}\ \bibnamefont {Odintsov}}, \bibinfo {author} {\bibfnamefont {V.~K.}\ \bibnamefont {Oikonomou}}, \bibinfo {author} {\bibfnamefont {N.}~\bibnamefont {Chatzarakis}},\ and\ \bibinfo {author} {\bibfnamefont {T.}~\bibnamefont {Paul}},\ }\bibfield  {title} {\bibinfo {title} {{Viable inflationary models in a ghost-free Gauss\textendash{}Bonnet theory of gravity}},\ }\href {https://doi.org/10.1140/epjc/s10052-019-7080-1} {\bibfield  {journal} {\bibinfo  {journal} {Eur. Phys. J. C}\ }\textbf {\bibinfo {volume} {79}},\ \bibinfo {pages} {565} (\bibinfo {year} {2019})},\ \Eprint {https://arxiv.org/abs/1907.00403} {arXiv:1907.00403 [gr-qc]} \BibitemShut {NoStop}%
\bibitem [{\citenamefont {Ashrafzadeh}\ and\ \citenamefont {Karami}(2024)}]{Ashrafzadeh:2023ndt}%
  \BibitemOpen
  \bibfield  {author} {\bibinfo {author} {\bibfnamefont {A.}~\bibnamefont {Ashrafzadeh}}\ and\ \bibinfo {author} {\bibfnamefont {K.}~\bibnamefont {Karami}},\ }\bibfield  {title} {\bibinfo {title} {{Primordial Black Holes in Scalar Field Inflation Coupled to the Gauss\textendash{}Bonnet Term with Fractional Power-law Potentials}},\ }\href {https://doi.org/10.3847/1538-4357/ad293f} {\bibfield  {journal} {\bibinfo  {journal} {Astrophys. J.}\ }\textbf {\bibinfo {volume} {965}},\ \bibinfo {pages} {11} (\bibinfo {year} {2024})},\ \Eprint {https://arxiv.org/abs/2309.16356} {arXiv:2309.16356 [astro-ph.CO]} \BibitemShut {NoStop}%
\bibitem [{\citenamefont {Oikonomou}(2024)}]{Oikonomou:2024etl}%
  \BibitemOpen
  \bibfield  {author} {\bibinfo {author} {\bibfnamefont {V.~K.}\ \bibnamefont {Oikonomou}},\ }\bibfield  {title} {\bibinfo {title} {{Revisiting Einstein-Gauss-Bonnet theories after GW170817}},\ }\href {https://doi.org/10.1016/j.physletb.2024.138890} {\bibfield  {journal} {\bibinfo  {journal} {Phys. Lett. B}\ }\textbf {\bibinfo {volume} {856}},\ \bibinfo {pages} {138890} (\bibinfo {year} {2024})},\ \Eprint {https://arxiv.org/abs/2407.12155} {arXiv:2407.12155 [gr-qc]} \BibitemShut {NoStop}%
\bibitem [{\citenamefont {Oikonomou}\ \emph {et~al.}(2024)\citenamefont {Oikonomou}, \citenamefont {Tsyba},\ and\ \citenamefont {Razina}}]{Oikonomou:2024jqv}%
  \BibitemOpen
  \bibfield  {author} {\bibinfo {author} {\bibfnamefont {V.~K.}\ \bibnamefont {Oikonomou}}, \bibinfo {author} {\bibfnamefont {P.}~\bibnamefont {Tsyba}},\ and\ \bibinfo {author} {\bibfnamefont {O.}~\bibnamefont {Razina}},\ }\bibfield  {title} {\bibinfo {title} {{Einstein\textendash{}Gauss\textendash{}Bonnet cosmological theories at reheating and at the end of the inflationary era}},\ }\href {https://doi.org/10.1016/j.aop.2024.169597} {\bibfield  {journal} {\bibinfo  {journal} {Annals Phys.}\ }\textbf {\bibinfo {volume} {462}},\ \bibinfo {pages} {169597} (\bibinfo {year} {2024})},\ \Eprint {https://arxiv.org/abs/2401.11273} {arXiv:2401.11273 [gr-qc]} \BibitemShut {NoStop}%
\bibitem [{\citenamefont {Odintsov}\ \emph {et~al.}(2023{\natexlab{a}})\citenamefont {Odintsov}, \citenamefont {Oikonomou}, \citenamefont {Giannakoudi}, \citenamefont {Fronimos},\ and\ \citenamefont {Lymperiadou}}]{Odintsov:2023weg}%
  \BibitemOpen
  \bibfield  {author} {\bibinfo {author} {\bibfnamefont {S.~D.}\ \bibnamefont {Odintsov}}, \bibinfo {author} {\bibfnamefont {V.~K.}\ \bibnamefont {Oikonomou}}, \bibinfo {author} {\bibfnamefont {I.}~\bibnamefont {Giannakoudi}}, \bibinfo {author} {\bibfnamefont {F.~P.}\ \bibnamefont {Fronimos}},\ and\ \bibinfo {author} {\bibfnamefont {E.~C.}\ \bibnamefont {Lymperiadou}},\ }\bibfield  {title} {\bibinfo {title} {{Recent Advances in Inflation}},\ }\href {https://doi.org/10.3390/sym15091701} {\bibfield  {journal} {\bibinfo  {journal} {Symmetry}\ }\textbf {\bibinfo {volume} {15}},\ \bibinfo {pages} {1701} (\bibinfo {year} {2023}{\natexlab{a}})},\ \Eprint {https://arxiv.org/abs/2307.16308} {arXiv:2307.16308 [gr-qc]} \BibitemShut {NoStop}%
\bibitem [{\citenamefont {Kawai}\ and\ \citenamefont {Soda}(1999)}]{Kawai:1999pw}%
  \BibitemOpen
  \bibfield  {author} {\bibinfo {author} {\bibfnamefont {S.}~\bibnamefont {Kawai}}\ and\ \bibinfo {author} {\bibfnamefont {J.}~\bibnamefont {Soda}},\ }\bibfield  {title} {\bibinfo {title} {{Evolution of fluctuations during graceful exit in string cosmology}},\ }\href {https://doi.org/10.1016/S0370-2693(99)00736-4} {\bibfield  {journal} {\bibinfo  {journal} {Phys. Lett. B}\ }\textbf {\bibinfo {volume} {460}},\ \bibinfo {pages} {41} (\bibinfo {year} {1999})},\ \Eprint {https://arxiv.org/abs/gr-qc/9903017} {arXiv:gr-qc/9903017} \BibitemShut {NoStop}%
\bibitem [{\citenamefont {Kawai}\ \emph {et~al.}(1998)\citenamefont {Kawai}, \citenamefont {Sakagami},\ and\ \citenamefont {Soda}}]{Kawai:1998ab}%
  \BibitemOpen
  \bibfield  {author} {\bibinfo {author} {\bibfnamefont {S.}~\bibnamefont {Kawai}}, \bibinfo {author} {\bibfnamefont {M.-a.}\ \bibnamefont {Sakagami}},\ and\ \bibinfo {author} {\bibfnamefont {J.}~\bibnamefont {Soda}},\ }\bibfield  {title} {\bibinfo {title} {{Instability of one loop superstring cosmology}},\ }\href {https://doi.org/10.1016/S0370-2693(98)00925-3} {\bibfield  {journal} {\bibinfo  {journal} {Phys. Lett. B}\ }\textbf {\bibinfo {volume} {437}},\ \bibinfo {pages} {284} (\bibinfo {year} {1998})},\ \Eprint {https://arxiv.org/abs/gr-qc/9802033} {arXiv:gr-qc/9802033} \BibitemShut {NoStop}%
\bibitem [{\citenamefont {Nojiri}\ and\ \citenamefont {Odintsov}(2024{\natexlab{a}})}]{Nojiri:2024hau}%
  \BibitemOpen
  \bibfield  {author} {\bibinfo {author} {\bibfnamefont {S.}~\bibnamefont {Nojiri}}\ and\ \bibinfo {author} {\bibfnamefont {S.~D.}\ \bibnamefont {Odintsov}},\ }\bibfield  {title} {\bibinfo {title} {{$F(Q)$ gravity with Gauss-Bonnet corrections: from early-time inflation to late-time acceleration}},\ }\href@noop {} {\  (\bibinfo {year} {2024}{\natexlab{a}})},\ \Eprint {https://arxiv.org/abs/2406.12558} {arXiv:2406.12558 [gr-qc]} \BibitemShut {NoStop}%
\bibitem [{\citenamefont {Nojiri}\ and\ \citenamefont {Odintsov}(2024{\natexlab{b}})}]{Nojiri:2024zab}%
  \BibitemOpen
  \bibfield  {author} {\bibinfo {author} {\bibfnamefont {S.}~\bibnamefont {Nojiri}}\ and\ \bibinfo {author} {\bibfnamefont {S.~D.}\ \bibnamefont {Odintsov}},\ }\bibfield  {title} {\bibinfo {title} {{Well-defined f(Q) gravity, reconstruction of FLRW spacetime and unification of inflation with dark energy epoch}},\ }\href {https://doi.org/10.1016/j.dark.2024.101538} {\bibfield  {journal} {\bibinfo  {journal} {Phys. Dark Univ.}\ }\textbf {\bibinfo {volume} {45}},\ \bibinfo {pages} {101538} (\bibinfo {year} {2024}{\natexlab{b}})},\ \Eprint {https://arxiv.org/abs/2404.18427} {arXiv:2404.18427 [gr-qc]} \BibitemShut {NoStop}%
\bibitem [{\citenamefont {Elizalde}\ \emph {et~al.}(2024)\citenamefont {Elizalde}, \citenamefont {Nojiri}, \citenamefont {Odintsov},\ and\ \citenamefont {Oikonomou}}]{Elizalde:2023rds}%
  \BibitemOpen
  \bibfield  {author} {\bibinfo {author} {\bibfnamefont {E.}~\bibnamefont {Elizalde}}, \bibinfo {author} {\bibfnamefont {S.}~\bibnamefont {Nojiri}}, \bibinfo {author} {\bibfnamefont {S.~D.}\ \bibnamefont {Odintsov}},\ and\ \bibinfo {author} {\bibfnamefont {V.~K.}\ \bibnamefont {Oikonomou}},\ }\bibfield  {title} {\bibinfo {title} {{Propagation of gravitational waves in a dynamical wormhole background for two-scalar Einstein\textendash{}Gauss\textendash{}Bonnet theory}},\ }\href {https://doi.org/10.1016/j.dark.2024.101536} {\bibfield  {journal} {\bibinfo  {journal} {Phys. Dark Univ.}\ }\textbf {\bibinfo {volume} {45}},\ \bibinfo {pages} {101536} (\bibinfo {year} {2024})},\ \Eprint {https://arxiv.org/abs/2312.02889} {arXiv:2312.02889 [gr-qc]} \BibitemShut {NoStop}%
\bibitem [{\citenamefont {Nojiri}\ \emph {et~al.}(2023)\citenamefont {Nojiri}, \citenamefont {Odintsov},\ and\ \citenamefont {S\'aez-Chill\'on~G\'omez}}]{Nojiri:2023mvi}%
  \BibitemOpen
  \bibfield  {author} {\bibinfo {author} {\bibfnamefont {S.}~\bibnamefont {Nojiri}}, \bibinfo {author} {\bibfnamefont {S.~D.}\ \bibnamefont {Odintsov}},\ and\ \bibinfo {author} {\bibfnamefont {D.}~\bibnamefont {S\'aez-Chill\'on~G\'omez}},\ }\bibfield  {title} {\bibinfo {title} {{Unifying inflation with early and late dark energy in Einstein\textendash{}Gauss\textendash{}Bonnet gravity}},\ }\href {https://doi.org/10.1016/j.dark.2023.101238} {\bibfield  {journal} {\bibinfo  {journal} {Phys. Dark Univ.}\ }\textbf {\bibinfo {volume} {41}},\ \bibinfo {pages} {101238} (\bibinfo {year} {2023})},\ \Eprint {https://arxiv.org/abs/2304.08255} {arXiv:2304.08255 [gr-qc]} \BibitemShut {NoStop}%
\bibitem [{\citenamefont {Odintsov}\ \emph {et~al.}(2023{\natexlab{b}})\citenamefont {Odintsov}, \citenamefont {Oikonomou},\ and\ \citenamefont {Fronimos}}]{Odintsov:2023aaw}%
  \BibitemOpen
  \bibfield  {author} {\bibinfo {author} {\bibfnamefont {S.~D.}\ \bibnamefont {Odintsov}}, \bibinfo {author} {\bibfnamefont {V.~K.}\ \bibnamefont {Oikonomou}},\ and\ \bibinfo {author} {\bibfnamefont {F.~P.}\ \bibnamefont {Fronimos}},\ }\bibfield  {title} {\bibinfo {title} {{Inflationary Dynamics and Swampland Criteria for Modified Gauss-Bonnet Gravity Compatible with GW170817}},\ }\href {https://doi.org/10.1103/PhysRevD.107.084007} {\bibfield  {journal} {\bibinfo  {journal} {Phys. Rev. D}\ }\textbf {\bibinfo {volume} {107}},\ \bibinfo {pages} {08} (\bibinfo {year} {2023}{\natexlab{b}})},\ \Eprint {https://arxiv.org/abs/2303.14594} {arXiv:2303.14594 [gr-qc]} \BibitemShut {NoStop}%
\bibitem [{\citenamefont {Odintsov}\ and\ \citenamefont {Oikonomou}(2022{\natexlab{a}})}]{Odintsov:2022rok}%
  \BibitemOpen
  \bibfield  {author} {\bibinfo {author} {\bibfnamefont {S.~D.}\ \bibnamefont {Odintsov}}\ and\ \bibinfo {author} {\bibfnamefont {V.~K.}\ \bibnamefont {Oikonomou}},\ }\bibfield  {title} {\bibinfo {title} {{Running of the spectral index and inflationary dynamics of F(R) gravity}},\ }\href {https://doi.org/10.1016/j.physletb.2022.137353} {\bibfield  {journal} {\bibinfo  {journal} {Phys. Lett. B}\ }\textbf {\bibinfo {volume} {833}},\ \bibinfo {pages} {137353} (\bibinfo {year} {2022}{\natexlab{a}})},\ \Eprint {https://arxiv.org/abs/2206.06024} {arXiv:2206.06024 [gr-qc]} \BibitemShut {NoStop}%
\bibitem [{\citenamefont {Odintsov}\ and\ \citenamefont {Oikonomou}(2022{\natexlab{b}})}]{Odintsov:2021urx}%
  \BibitemOpen
  \bibfield  {author} {\bibinfo {author} {\bibfnamefont {S.~D.}\ \bibnamefont {Odintsov}}\ and\ \bibinfo {author} {\bibfnamefont {V.~K.}\ \bibnamefont {Oikonomou}},\ }\bibfield  {title} {\bibinfo {title} {{Pre-inflationary bounce effects on primordial gravitational waves of f(R) gravity}},\ }\href {https://doi.org/10.1016/j.physletb.2021.136817} {\bibfield  {journal} {\bibinfo  {journal} {Phys. Lett. B}\ }\textbf {\bibinfo {volume} {824}},\ \bibinfo {pages} {136817} (\bibinfo {year} {2022}{\natexlab{b}})},\ \Eprint {https://arxiv.org/abs/2112.02584} {arXiv:2112.02584 [gr-qc]} \BibitemShut {NoStop}%
\bibitem [{\citenamefont {Kawai}\ and\ \citenamefont {Kim}(2023)}]{Kawai:2023nqs}%
  \BibitemOpen
  \bibfield  {author} {\bibinfo {author} {\bibfnamefont {S.}~\bibnamefont {Kawai}}\ and\ \bibinfo {author} {\bibfnamefont {J.}~\bibnamefont {Kim}},\ }\bibfield  {title} {\bibinfo {title} {{Probing the inflationary moduli space with gravitational waves}},\ }\href {https://doi.org/10.1103/PhysRevD.108.103537} {\bibfield  {journal} {\bibinfo  {journal} {Phys. Rev. D}\ }\textbf {\bibinfo {volume} {108}},\ \bibinfo {pages} {103537} (\bibinfo {year} {2023})},\ \Eprint {https://arxiv.org/abs/2308.13272} {arXiv:2308.13272 [astro-ph.CO]} \BibitemShut {NoStop}%
\bibitem [{\citenamefont {Mudrunka}\ and\ \citenamefont {Nakayama}(2025)}]{Mudrunka:2025xcg}%
  \BibitemOpen
  \bibfield  {author} {\bibinfo {author} {\bibfnamefont {K.}~\bibnamefont {Mudrunka}}\ and\ \bibinfo {author} {\bibfnamefont {K.}~\bibnamefont {Nakayama}},\ }\bibfield  {title} {\bibinfo {title} {{Inflation with Gauss-Bonnet correction: beyond slow-roll}},\ }\href@noop {} {\  (\bibinfo {year} {2025})},\ \Eprint {https://arxiv.org/abs/2504.01365} {arXiv:2504.01365 [gr-qc]} \BibitemShut {NoStop}%
\bibitem [{\citenamefont {Kawai}\ and\ \citenamefont {Kim}(2021{\natexlab{b}})}]{Kawai:2021edk}%
  \BibitemOpen
  \bibfield  {author} {\bibinfo {author} {\bibfnamefont {S.}~\bibnamefont {Kawai}}\ and\ \bibinfo {author} {\bibfnamefont {J.}~\bibnamefont {Kim}},\ }\bibfield  {title} {\bibinfo {title} {{Primordial black holes from Gauss-Bonnet-corrected single field inflation}},\ }\href {https://doi.org/10.1103/PhysRevD.104.083545} {\bibfield  {journal} {\bibinfo  {journal} {Phys. Rev. D}\ }\textbf {\bibinfo {volume} {104}},\ \bibinfo {pages} {083545} (\bibinfo {year} {2021}{\natexlab{b}})},\ \Eprint {https://arxiv.org/abs/2108.01340} {arXiv:2108.01340 [astro-ph.CO]} \BibitemShut {NoStop}%
\bibitem [{\citenamefont {Yogesh}\ and\ \citenamefont {Mohammadi}(2025)}]{Yogesh:2025hll}%
  \BibitemOpen
  \bibfield  {author} {\bibinfo {author} {\bibnamefont {Yogesh}}\ and\ \bibinfo {author} {\bibfnamefont {A.}~\bibnamefont {Mohammadi}},\ }\bibfield  {title} {\bibinfo {title} {{Non-Standard Thermal History and Formation of Primordial Black Holes in Einstein-Gauss-Bonnet Gravity}},\ }\href@noop {} {\  (\bibinfo {year} {2025})},\ \Eprint {https://arxiv.org/abs/2501.01867} {arXiv:2501.01867 [gr-qc]} \BibitemShut {NoStop}%
\bibitem [{\citenamefont {Yi}\ \emph {et~al.}(2018)\citenamefont {Yi}, \citenamefont {Gong},\ and\ \citenamefont {Sabir}}]{Yi:2018gse}%
  \BibitemOpen
  \bibfield  {author} {\bibinfo {author} {\bibfnamefont {Z.}~\bibnamefont {Yi}}, \bibinfo {author} {\bibfnamefont {Y.}~\bibnamefont {Gong}},\ and\ \bibinfo {author} {\bibfnamefont {M.}~\bibnamefont {Sabir}},\ }\bibfield  {title} {\bibinfo {title} {{Inflation with Gauss-Bonnet coupling}},\ }\href {https://doi.org/10.1103/PhysRevD.98.083521} {\bibfield  {journal} {\bibinfo  {journal} {Phys. Rev. D}\ }\textbf {\bibinfo {volume} {98}},\ \bibinfo {pages} {083521} (\bibinfo {year} {2018})},\ \Eprint {https://arxiv.org/abs/1804.09116} {arXiv:1804.09116 [gr-qc]} \BibitemShut {NoStop}%
\bibitem [{\citenamefont {Rashidi}\ and\ \citenamefont {Nozari}(2020)}]{Rashidi:2020wwg}%
  \BibitemOpen
  \bibfield  {author} {\bibinfo {author} {\bibfnamefont {N.}~\bibnamefont {Rashidi}}\ and\ \bibinfo {author} {\bibfnamefont {K.}~\bibnamefont {Nozari}},\ }\bibfield  {title} {\bibinfo {title} {{Gauss-Bonnet Inflation after Planck2018}},\ }\href {https://doi.org/10.3847/1538-4357/ab6a10} {\bibfield  {journal} {\bibinfo  {journal} {Astrophys. J.}\ }\textbf {\bibinfo {volume} {890}},\ \bibinfo {pages} {58} (\bibinfo {year} {2020})},\ \Eprint {https://arxiv.org/abs/2001.07012} {arXiv:2001.07012 [astro-ph.CO]} \BibitemShut {NoStop}%
\bibitem [{\citenamefont {Starobinsky}(1980)}]{Starobinsky:1980te}%
  \BibitemOpen
  \bibfield  {author} {\bibinfo {author} {\bibfnamefont {A.~A.}\ \bibnamefont {Starobinsky}},\ }\bibfield  {title} {\bibinfo {title} {{A New Type of Isotropic Cosmological Models Without Singularity}},\ }\href {https://doi.org/10.1016/0370-2693(80)90670-X} {\bibfield  {journal} {\bibinfo  {journal} {Phys. Lett. B}\ }\textbf {\bibinfo {volume} {91}},\ \bibinfo {pages} {99} (\bibinfo {year} {1980})}\BibitemShut {NoStop}%
\bibitem [{\citenamefont {Pozdeeva}\ \emph {et~al.}(2024)\citenamefont {Pozdeeva}, \citenamefont {Skugoreva}, \citenamefont {Toporensky},\ and\ \citenamefont {Vernov}}]{Pozdeeva:2024ihc}%
  \BibitemOpen
  \bibfield  {author} {\bibinfo {author} {\bibfnamefont {E.~O.}\ \bibnamefont {Pozdeeva}}, \bibinfo {author} {\bibfnamefont {M.~A.}\ \bibnamefont {Skugoreva}}, \bibinfo {author} {\bibfnamefont {A.~V.}\ \bibnamefont {Toporensky}},\ and\ \bibinfo {author} {\bibfnamefont {S.~Y.}\ \bibnamefont {Vernov}},\ }\bibfield  {title} {\bibinfo {title} {{New slow-roll approximations for inflation in Einstein-Gauss-Bonnet gravity}},\ }\href {https://doi.org/10.1088/1475-7516/2024/09/050} {\bibfield  {journal} {\bibinfo  {journal} {JCAP}\ }\textbf {\bibinfo {volume} {09}},\ \bibinfo {pages} {050}},\ \Eprint {https://arxiv.org/abs/2403.06147} {arXiv:2403.06147 [gr-qc]} \BibitemShut {NoStop}%
\bibitem [{\citenamefont {Khan}\ and\ \citenamefont {Yogesh}(2022)}]{Khan:2022odn}%
  \BibitemOpen
  \bibfield  {author} {\bibinfo {author} {\bibfnamefont {H.~A.}\ \bibnamefont {Khan}}\ and\ \bibinfo {author} {\bibnamefont {Yogesh}},\ }\bibfield  {title} {\bibinfo {title} {{Study of Goldstone inflation in the domain of Einstein-Gauss-Bonnet gravity}},\ }\href {https://doi.org/10.1103/PhysRevD.105.063526} {\bibfield  {journal} {\bibinfo  {journal} {Phys. Rev. D}\ }\textbf {\bibinfo {volume} {105}},\ \bibinfo {pages} {063526} (\bibinfo {year} {2022})},\ \Eprint {https://arxiv.org/abs/2201.06439} {arXiv:2201.06439 [astro-ph.CO]} \BibitemShut {NoStop}%
\bibitem [{\citenamefont {Gangopadhyay}\ \emph {et~al.}(2023)\citenamefont {Gangopadhyay}, \citenamefont {Khan},\ and\ \citenamefont {Yogesh}}]{Gangopadhyay:2022vgh}%
  \BibitemOpen
  \bibfield  {author} {\bibinfo {author} {\bibfnamefont {M.~R.}\ \bibnamefont {Gangopadhyay}}, \bibinfo {author} {\bibfnamefont {H.~A.}\ \bibnamefont {Khan}},\ and\ \bibinfo {author} {\bibnamefont {Yogesh}},\ }\bibfield  {title} {\bibinfo {title} {{A case study of small field inflationary dynamics in the Einstein\textendash{}Gauss\textendash{}Bonnet framework in the light of GW170817}},\ }\href {https://doi.org/10.1016/j.dark.2023.101177} {\bibfield  {journal} {\bibinfo  {journal} {Phys. Dark Univ.}\ }\textbf {\bibinfo {volume} {40}},\ \bibinfo {pages} {101177} (\bibinfo {year} {2023})},\ \Eprint {https://arxiv.org/abs/2205.15261} {arXiv:2205.15261 [astro-ph.CO]} \BibitemShut {NoStop}%
\bibitem [{\citenamefont {Yogesh}\ \emph {et~al.}(2024)\citenamefont {Yogesh}, \citenamefont {Bhat},\ and\ \citenamefont {Gangopadhyay}}]{Yogesh:2024mpa}%
  \BibitemOpen
  \bibfield  {author} {\bibinfo {author} {\bibnamefont {Yogesh}}, \bibinfo {author} {\bibfnamefont {I.~A.}\ \bibnamefont {Bhat}},\ and\ \bibinfo {author} {\bibfnamefont {M.~R.}\ \bibnamefont {Gangopadhyay}},\ }\bibfield  {title} {\bibinfo {title} {{Inflationary Dynamics in Einstein-Gauss-Bonnet Gravity Using New Slow-Roll Approximations considering Generalised Reheating}},\ }\href@noop {} {\  (\bibinfo {year} {2024})},\ \Eprint {https://arxiv.org/abs/2408.01670} {arXiv:2408.01670 [astro-ph.CO]} \BibitemShut {NoStop}%
\bibitem [{\citenamefont {Pozdeeva}\ \emph {et~al.}(2019)\citenamefont {Pozdeeva}, \citenamefont {Sami}, \citenamefont {Toporensky},\ and\ \citenamefont {Vernov}}]{Pozdeeva:2019agu}%
  \BibitemOpen
  \bibfield  {author} {\bibinfo {author} {\bibfnamefont {E.~O.}\ \bibnamefont {Pozdeeva}}, \bibinfo {author} {\bibfnamefont {M.}~\bibnamefont {Sami}}, \bibinfo {author} {\bibfnamefont {A.~V.}\ \bibnamefont {Toporensky}},\ and\ \bibinfo {author} {\bibfnamefont {S.~Y.}\ \bibnamefont {Vernov}},\ }\bibfield  {title} {\bibinfo {title} {{Stability analysis of de Sitter solutions in models with the Gauss-Bonnet term}},\ }\href {https://doi.org/10.1103/PhysRevD.100.083527} {\bibfield  {journal} {\bibinfo  {journal} {Phys. Rev. D}\ }\textbf {\bibinfo {volume} {100}},\ \bibinfo {pages} {083527} (\bibinfo {year} {2019})},\ \Eprint {https://arxiv.org/abs/1905.05085} {arXiv:1905.05085 [gr-qc]} \BibitemShut {NoStop}%
\bibitem [{\citenamefont {Vernov}\ and\ \citenamefont {Pozdeeva}(2021)}]{Vernov:2021hxo}%
  \BibitemOpen
  \bibfield  {author} {\bibinfo {author} {\bibfnamefont {S.}~\bibnamefont {Vernov}}\ and\ \bibinfo {author} {\bibfnamefont {E.}~\bibnamefont {Pozdeeva}},\ }\bibfield  {title} {\bibinfo {title} {{De Sitter Solutions in Einstein\textendash{}Gauss\textendash{}Bonnet Gravity}},\ }\href {https://doi.org/10.3390/universe7050149} {\bibfield  {journal} {\bibinfo  {journal} {Universe}\ }\textbf {\bibinfo {volume} {7}},\ \bibinfo {pages} {149} (\bibinfo {year} {2021})},\ \Eprint {https://arxiv.org/abs/2104.11111} {arXiv:2104.11111 [gr-qc]} \BibitemShut {NoStop}%
\bibitem [{\citenamefont {Cook}\ \emph {et~al.}(2015)\citenamefont {Cook}, \citenamefont {Dimastrogiovanni}, \citenamefont {Easson},\ and\ \citenamefont {Krauss}}]{Cook:2015vqa}%
  \BibitemOpen
  \bibfield  {author} {\bibinfo {author} {\bibfnamefont {J.~L.}\ \bibnamefont {Cook}}, \bibinfo {author} {\bibfnamefont {E.}~\bibnamefont {Dimastrogiovanni}}, \bibinfo {author} {\bibfnamefont {D.~A.}\ \bibnamefont {Easson}},\ and\ \bibinfo {author} {\bibfnamefont {L.~M.}\ \bibnamefont {Krauss}},\ }\bibfield  {title} {\bibinfo {title} {{Reheating predictions in single field inflation}},\ }\href {https://doi.org/10.1088/1475-7516/2015/04/047} {\bibfield  {journal} {\bibinfo  {journal} {JCAP}\ }\textbf {\bibinfo {volume} {04}},\ \bibinfo {pages} {047}},\ \Eprint {https://arxiv.org/abs/1502.04673} {arXiv:1502.04673 [astro-ph.CO]} \BibitemShut {NoStop}%
\bibitem [{\citenamefont {Adhikari}\ \emph {et~al.}(2022)\citenamefont {Adhikari}, \citenamefont {Gangopadhyay},\ and\ \citenamefont {Yogesh}}]{Adhikari:2019uaw}%
  \BibitemOpen
  \bibfield  {author} {\bibinfo {author} {\bibfnamefont {R.}~\bibnamefont {Adhikari}}, \bibinfo {author} {\bibfnamefont {M.~R.}\ \bibnamefont {Gangopadhyay}},\ and\ \bibinfo {author} {\bibnamefont {Yogesh}},\ }\bibfield  {title} {\bibinfo {title} {{Lower Tensor to Scalar Ratio in a SUGRA Motivated Inflationary Potential}},\ }\href {https://doi.org/10.1134/S0202289322010029} {\bibfield  {journal} {\bibinfo  {journal} {Grav. Cosmol.}\ }\textbf {\bibinfo {volume} {28}},\ \bibinfo {pages} {1} (\bibinfo {year} {2022})},\ \Eprint {https://arxiv.org/abs/1909.07217} {arXiv:1909.07217 [astro-ph.CO]} \BibitemShut {NoStop}%
\end{thebibliography}%

\end{document}